\documentclass{sig-alternate}

\usepackage{amsmath}
\usepackage{amssymb}
\usepackage{times}
\usepackage{url}
\usepackage{graphicx}
\usepackage{multirow}
\usepackage{ifthen}
\usepackage{supertabular,booktabs}
\newcommand{\doctype}{Extended}

\newcommand{\YX}[1]	{\textbf{[YannisX Note: {#1}]}}

\newcommand{\hide}[1]{}

\newcommand{\res}     {\ensuremath{\mathsf{rdfs:resource}}}
\newcommand{\class}   {\ensuremath{\mathsf{rdfs:class}}}
\newcommand{\prop}    {\ensuremath{\mathsf{rdf:property}}}

\newcommand{\type}    {\ensuremath{\mathsf{rdf:type}}}
\newcommand{\cisa}    {\ensuremath{\mathsf{rdfs:subClassOf}}}
\newcommand{\pisa}    {\ensuremath{\mathsf{rdfs:subPropertyOf}}}
\newcommand{\pdomain} {\ensuremath{\mathsf{rdfs:domain}}}
\newcommand{\prange}  {\ensuremath{\mathsf{rdfs:range}}}
\newcommand{\comment} {\ensuremath{\mathsf{rdfs:comment}}}
\newcommand{\labell}  {\ensuremath{\mathsf{rdfs:label}}}

\newcommand{\uris}		{\ensuremath{\mathbb{U}}}	
\newcommand{\literals}	{\ensuremath{\mathbb{L}}} 	
\newcommand{\triples}	{\ensuremath{\mathbb{T}}} 	
\newcommand{\tps}		{\ensuremath{\mathbb{TP}}}	
\newcommand{\vars}		{\ensuremath{\mathbb{V}}}	
\newcommand{\ds}			{\ensuremath{{\cal D}}}  		
\newcommand{\lang}		{\ensuremath{{\cal L}}}  		
\newcommand{\vold}		{\ensuremath{{D_{old}}}}	
\newcommand{\vnew}		{\ensuremath{{D_{new}}}}	

\newcommand{\chng}		{\ensuremath{\delta}}
\newcommand{\deltap}	{\ensuremath{\delta^+}} 	
\newcommand{\deltam}	{\ensuremath{\delta^-}} 	
\newcommand{\deltas}	{\ensuremath{\delta^s}} 	
\newcommand{\cond}		{\ensuremath{\phi}}		

\newcommand{\assoc}		{\ensuremath{\alpha}} 	
\newcommand{\assocset}	{\ensuremath{{\cal A}}} 	

\newcommand{\ExtOrSub}[2]{%
\ifthenelse{\equal{\doctype}{Extended}}
{#1}
{#2}
}

\newtheorem{definition}{Definition}  

\newcommand{\uu}[1]		{\ensuremath{#1}}
\newcommand{\triple}[3]{(\ensuremath{#1}, \ensuremath{#2}, \ensuremath{#3})} 
\newcommand{\op}[1] {\textit{#1}}

\newcommand{\isDimension}[1]{\ensuremath{\triple{#1}{\type}{\uu{qb:DimensionProperty}}}}
\newcommand{\isCodelist}[1]{\ensuremath{\triple{#1}{\type}{\uu{skos:ConceptScheme}}}}
\newcommand{\isObservation}[1]{\ensuremath{\triple{#1}{\type}{\uu{qb:Observation}}}}
\newcommand{\isMeasure}[1]{\ensuremath{\triple{#1}{\type}{\uu{qb:MeasureProperty}}}}
\newcommand{\isInstance}[1]{\ensuremath{\triple{#1}{\type}{\uu{skos:Concept}}}}
\newcommand{\isHierarchy}[1]{\ensuremath{\triple{#1}{\type}{\uu{qb:HierarchicalCodeList}}}}
\newcommand{\isFT}[1]{\ensuremath{\triple{#1}{\type}{\uu{qb:DataStructureDefinition}}}}
\newcommand{\isAttribute}[1]{\ensuremath{\triple{#1}{\type}{\uu{qb:AttributeProperty}}}}

\newcommand{\hasRange}[2]{\ensuremath{\triple{#1}{\mathsf{rdfs:range}}{#2}}}
\newcommand{\hasAttribute}[2]{\ensuremath{\triple{#1}{\uu{qb:attribute}}{#2}}}
\newcommand{\hasDataset}[2]{\ensuremath{\triple{#1}{\uu{qb:dataSet}}{#2}}}
\newcommand{\hasStructure}[2]{\ensuremath{\triple{#1}{\uu{qb:structure}}{#2}}}
\newcommand{\hasComponent}[2]{\ensuremath{\triple{#1}{\uu{qb:component}}{#2}}}
\newcommand{\hasComponentProperty}[2]{\ensuremath{\triple{#1}{\uu{qb:componentProperty}}{#2}}}
\newcommand{\hasMeasure}[2]{\ensuremath{\triple{#1}{\uu{qb:measure}}{#2}}}
\newcommand{\hasDimension}[2]{\ensuremath{\triple{#1}{\uu{qb:dimension}}{#2}}}
\newcommand{\hasCodelist}[2]{\ensuremath{\triple{#1}{\uu{qb:codeList}}{#2}}}
\newcommand{\inScheme}[2]{\ensuremath{\triple{#1}{\uu{skos:inScheme}}{#2}}}
\newcommand{\hasParent}[2]{\ensuremath{\triple{#1}{\uu{skos:broaderTransitive}}{#2}}}

\begin{document}

\title{A Flexible Framework for Defining, Representing and Detecting Changes on the Data Web}

\numberofauthors{5}
\author{
\alignauthor
Yannis Roussakis\\
       \affaddr{FORTH-ICS}\\
       \email{rousakis@ics.forth.gr}
\alignauthor
Ioannis Chrysakis\\
       \affaddr{FORTH-ICS}\\
       \email{hrysakis@ics.forth.gr}
\alignauthor 
Kostas Stefanidis\\
       \affaddr{FORTH-ICS}\\
       \email{kstef@ics.forth.gr}
\and
\alignauthor 
Giorgos Flouris\\
       \affaddr{FORTH-ICS}\\
       \email{fgeo@ics.forth.gr}
\alignauthor 
Yannis Stavrakas\\
       \affaddr{ATHENA-IMIS}\\
       \email{yannis@imis.athena-innovation.gr}
}

\maketitle

\begin{abstract}
The dynamic nature of Web data gives rise to a multitude of problems related to the identification, computation and management of the evolving versions and the related changes. In this paper, we consider the problem of change recognition in RDF datasets, i.e., the problem of identifying, and when possible give semantics to, the changes that led from one version of an RDF dataset to another. Despite our RDF focus, our approach is sufficiently general to engulf different data models that can be encoded in RDF, such as relational or multi-dimensional. In fact, we propose a flexible, extendible and data-model-independent methodology of defining changes that can capture the peculiarities and needs of different data models and applications, while being formally robust due to the satisfaction of the properties of completeness and unambiguity. Further, we propose an ontology of changes for storing the detected changes that allows automated processing and analysis of changes, cross-snapshot queries (spanning across different versions), as well as queries involving both changes and data. To detect changes and populate said ontology, we propose a customizable detection algorithm, which is applicable to different data models and applications requiring the detection of custom, user-defined changes. Finally, we provide a proof-of-concept application and evaluation of our framework for different data models.
\end{abstract}





\section{Introduction}
\label{sec:intro}

With the growing complexity of the WWW, we face a completely different way for creating, disseminating and consuming big volumes of information. Large-scale corporate, government, or even user-generated data are published and become available to a wide spectrum of users. DBpedia, Freebase, YAGO and Atlas are, among many others, examples of large data repositories, which store information about various entities, including their relationships. Typically, data in such datasets are represented using the RDF model~\cite{rdf}, in which information is stored in triples of the form (\textit{subject}, \textit{predicate}, \textit{object}), meaning that \textit{subject} is related to \textit{object} via \textit{predicate}. 

Dynamicity is an indispensable part of the current web; datasets are constantly evolving for several reasons, such as the inclusion of new experimental evidence or observations, or the correction of erroneous conceptualizations~\cite{freqevolution}. As an example, consider the detected number of changes between the versions 12.07 and 13.05, and 13.05 and 13.07 of Atlas, which are 879.5M and 801.2M triples, respectively, while between the versions 3.7 and 3.8, and 3.8 and 3.9 of the English DBpedia are 20.7M and 9.3M triples (Figure~\ref{fig:results_changes}). 

This constant evolution poses several research problems, which are related to the identification, computation, storage and management of the evolving versions. 
An important question is how to support complex changes, whose constituent changes are seemingly unrelated and may occur on disparate pieces of data, but together as a whole they have a semantically coherent meaning for an application domain. 
Clearly, understanding and detecting the changes of evolving datasets is a prerequisite to address these problems. In particular, finding the differences (deltas) between datasets has been proved to play a crucial role in various curation tasks, such as the synchronization of autonomously developed datasets versions~\cite{DBLP:Cloran2005}, or the visualization of the evolution history of a dataset~\cite{DBLP:Noy2006}. 
Deltas are also necessary in certain applications that require access to previous versions of a dataset to support historical or cross-snapshot queries (e.g.,~\cite{DBLP:conf/er/StefanidisCF14}), in order, for example, to identify past states of the dataset, understand the evolution process, or detect the source of errors in the current modelling.

In general, there are two ways to record the occurring changes. The first is by constantly \textit{monitoring} the dataset and logging any change, which requires a closed and controlled system, where all changes pass through a dedicated application that records every change as it happens. 
A more flexible method includes \textit{detecting} a posteriori the changes that happened. This avoids the need to have a controlled system, and allows remote users of a dataset to identify changes, even if they have no access to the actual change process or knowledge that it happened. 
Here, we focus on the latter case, which is more challenging and interesting, even though most of the proposed solutions (namely, the definition of a language of changes and the representation of the changes) are applicable in both scenarios.

Recognition of changes is based on 3 main pillars, namely \textit{defining}, \textit{representing} and \textit{detecting} changes. 
The first pillar (defining changes) is related to the definition of a \textit{language of changes}, which is based on the set of changes that are understandable by the system. Defining the language of changes includes identifying the types of changes that will be detected, their formal definition (e.g., their semantics and parameters), and takes into account the need for complex, domain-specific changes. It has been argued that the language of changes as a whole, should be well-behaved in the sense of satisfying certain properties, namely being \textit{complete and unambiguous}, so as to allow the generation of unique deltas in a deterministic manner~\cite{DBLP:journals/tods/PapavasileiouFFKC13}.
The definition of changes is given in Section~\ref{sec:define}.

The second pillar (representing changes) is related to the representation scheme used for storing the detected changes. This is necessary to allow a persistent representation and storage of the detected changes, as well as to support navigation among versions, analysis of the deltas, cross-snapshot or historic queries, and the raising of changes as first class citizens in a multi-version repository. Our approach is based on the definition of an \textit{ontology of changes} that satisfies these goals.
The representation scheme of the proposed changes is given in Section~\ref{sec:represent}.

The third pillar (detecting changes) is related to the algorithmic definition of the detection process. This process identifies the changes applied between any two versions, based on the actual change definitions in the language of changes, and stores the results in the ontology of changes. Our approach for change detection is based on the execution of appropriately defined SPARQL queries, whose answers determine the detected changes.
Details on the detection process are given in Section~\ref{sec:detect}.

A major challenge related to change detection and deltas is that no language of changes is suitable for all different applications and data models.
There are two reasons for that. First, the different types of data available on the Web are often represented using different models, which call for different changes.
Second, different uses (or users) of the data may require a different set of changes being reported, or the definition of special types of changes that happen often or are important for the operation of the data-driven application. 
Therefore, there is no one-size-fits-all solution for the problem of change recognition.

In this work, we propose a flexible framework for change recognition that can be applied to different data models and applications. Our intention is not to provide a single solution, but a generic methodology for defining the three components of a multi-version repository, through which different solutions, suitable for different needs, can be developed. 
Even though our framework is designed for RDF, our approach is applicable to any data model representable in RDF, i.e., RDF is used as a unifying underlying model to achieve this generality. 

In overall, our contributions are as follows:
\begin{itemize}
\item Regarding the definition of changes, the objective of generality is met by providing two different types of changes, namely \textit{simple} and \textit{complex} (see Section~\ref{sec:define}). The former type is meant to capture fine-grained changes for the data model at hand, and should meet the requirements of completeness and unambiguity.
The latter type is meant to capture more coarse-grained, or specialized, changes that are useful for the application at hand; this allows a customized behaviour of the change detection process, depending on the actual needs of the application. Complex changes are totally dynamic, and can be defined even at run-time, greatly enhancing the flexibility of our approach. 
\item For the purposes of representation, the proposed ontology of changes is designed to be generic enough, so as to be customizable for different languages of changes.
This allows us to meet the aforementioned generality goal, as detailed in Section~\ref{sec:represent}. 
\item The detection process is defined in a customizable manner, as the core detection algorithm is agnostic to the set of simple or complex changes used. The customization for the actual language of changes employed is based on the definition of appropriate SPARQL queries (one per change) that come as parameters to the algorithm, thereby allowing new changes to be easily defined. Details are given in Section~\ref{sec:detect}.
\item An additional contribution is the application of this work for data models other than the pure RDF model, namely the multi-dimensional model that has been transformed to the RDF Data Cube vocabulary, while it could be also applied similarly to the relational model. This application is meant as a proof-of-concept that the proposed approach is indeed capable of supporting diverse needs, and is not exhaustive in terms of the types of applications or models that our framework supports.
\item Finally, we provide experiments showing that the most crucial parameter in the change recognition process is the total number of detected changes rather than the size of the examined datasets. We prove also that the number of simple changes is proportional to the number of triples which will be inserted into the ontology. Except from the evaluation aspect of our framework, our experiments offer a detailed look on the evolution of real-world, big datasets, as we record changes and provide an analysis of their form. 
\end{itemize}

The approach proposed in this paper extends our previous work on change detection~\cite{DBLP:journals/tods/PapavasileiouFFKC13}, by providing a more generic change definition framework; here, we experience significantly improved performance (about 1 order of magnitude), scalability, as well as increased generality and applicability.




\section{Preliminaries}
\label{sec:prelim}

We consider two disjoint sets \uris, \literals, denoting the {\em URIs} and {\em literals} (we ignore here blank nodes that  can be avoided when data are published according to the Linked Data paradigm). An {\em RDF triple}~\cite{rdf} is a tuple of the form {\em (subject, predicate, object)} and asserts the fact that {\em subject} is associated with {\em object} through {\em predicate}. The set 
\triples $=$ $\uris \times \uris \times (\uris \cup \literals)$ 
is the set of all RDF triples.  An {\em RDF dataset} \ds\
consists of a set of RDF triples. 
In the following, we denote by \vold, \vnew\ the old and new versions, respectively, of a dataset \ds.

SPARQL 1.1~\cite{w3c-sparql-query} is the official
W3C recommendation language for querying RDF graphs.
The building block of a SPARQL statement is a {\em triple pattern}
$tp$ that is like an RDF triple, but may have a \emph{variable} (prefixed with character $?$) in any
of its \emph{subject}, \emph{predicate}, or \emph{object} positions; variables are taken from an infinite set of variables \vars, disjoint from the sets \uris, \literals, so the set of triple patterns is: 
\tps $=$ $(\uris \cup \vars) \times (\uris \cup \vars) \times (\uris \cup \literals \cup \vars)$.
SPARQL triple patterns can be combined into {\em graph patterns} $gp$, using operators like
\emph{join} ({\sf ``.''}), 
\emph{optional} ({\sf OPTIONAL}) and 
\emph{union} ({\sf UNION})~\cite{sparql-semantics2} and may also include
\emph{conditions} (using {\sf FILTER}). 
In this work, we are only interested in SELECT SPARQL queries, which are of the form:
``$SELECT \; v_1,\dots,v_n \; WHERE \; gp$'', where $n>0$, $v_i \in \vars$ and $gp$ is  a graph pattern.

For the evaluation of SPARQL queries, we follow the semantics discussed in~\cite{sparql-semantics,sparql-semantics2}. 
Evaluation is based on \textit{mappings}, which are partial functions 
$\mu: \vars \mapsto \uris \cup \literals$ that associate variables with URIs or literals (abusing notation, 
$\mu(tp)$ is used to denote the result of replacing the variables in $tp$ with their assigned values according to $\mu$). 
Then, the evaluation of a SPARQL triple pattern $tp$ on a dataset \ds\ returns a set of mappings (denoted by $[[tp]]^{\ds}$) such that $\mu(tp) \in \ds$ for $\mu \in [[tp]]^{\ds}$. 
This idea is extended to graph patterns by considering the semantics of the various operators  (e.g., $[[tp_1 \; UNION \; tp_2]]^{\ds} = [[tp_1]]^{\ds} \cup [[tp_2]]^{\ds}$).
Given a SPARQL query ``$SELECT \; v_1,\dots,v_n \; WHERE \; gp$'', its result when applied on \ds\ is $(\mu(v_1),\dots,\mu(v_n))$ for $\mu \in [[gp]]^{\ds}$.
For the precise semantics and further details on the evaluation of SPARQL queries, the reader is referred to~\cite{sparql-semantics,sparql-semantics2}.

\section{Defining Changes}
\label{sec:define}

Our purpose in this work is to provide a change recognition method, which, given two (subsequent) dataset versions \vold, \vnew, would produce their \emph{delta} ($\Delta$), i.e., a formal description of the changes that were made to get from \vold\ to \vnew. A delta is based on a \textit{language of changes} (\lang), i.e., a set of formal definitions of the changes that the delta could contain and, subsequently, the change detection method understands and detects; these changes 
should correspond to the {\em evolution primitives} of the data model under consideration. 

A language of changes, in its simplest form, consists of additions and deletions of elements (i.e., triples for the RDF case) from a dataset, i.e., changes of the form $Add(t)/Del(t)$.
Such a delta is called a \emph{low-level delta}~\cite{Zeginis:2011:CDR:1993053.1993056} and can be easily computed as follows:
$\Delta(\vold,\vnew) = \{Add(t) \mid t \in \vnew \setminus \vold \} \cup \{Del(t) \mid t \in \vold \setminus \vnew \}$.

Low-level languages (and deltas) are easy to define and detect, and have several nice properties~\cite{Zeginis:2011:CDR:1993053.1993056}; however, the representation of changes at the level of (added/deleted) triples, leads to a syntactic delta, which does not capture the semantics of a change and generates results that are not intuitive enough for the human user. 
For example, in the RDF context, the plain deletion of an individual (class instance) would correspond to a multitude of triple deletions (namely, all the triples that contain this URI, such as property instances); listing all these changes does not immediately convey the message that an individual was deleted, and the human observer may find it hard to decipher the intent of the actual changes that took place~\cite{DBLP:journals/tods/PapavasileiouFFKC13}.
This problem is even more serious for other data models (e.g., multi-dimensional) that are translated into RDF; in this case, the low-level deltas, which are described in RDF jargon, need to be ``translated back'' into the original data model, making deciphering even more difficult.


To address this problem, \emph{high-level deltas} have been proposed~\cite{DBLP:journals/tods/PapavasileiouFFKC13}, which aim to describe changes at a more intuitive level, in order to make them more human-understandable. In the above example, the output would be a ``delete individual'' change, that immediately conveys the message that all triples that include a given URI (individual) were deleted.
The main idea behind achieving this is to group low-level changes into high-level ones, under some conditions. 
For the purposes of this paper, we organize high-level changes in two major types, namely {\em simple} and {\em complex}, each of which represents changes with a certain granularity and role in the model. 
Details on these two types 
are given below.

\subsection{Simple Changes}
\label{subsec:simple}

In general, the role of \textit{simple changes} is to describe changes (evolution primitives) specific to the data model at hand; e.g., for the multi-dimensional model, we would expect changes like ``Add\_Di- mension'' or ``Attach\_Type\_To\_Measure''. 
Using simple changes, the user can abstract from the syntactical and representational peculiarities of the underlying data model (including its possible translation to RDF format), thereby making deltas more intuitive.

A simple change (e.g., Attach\_Type\_To\_Measure(m,t)), is composed of the \textit{change name} (i.e., Attach\_Type\_To\_Measure) and the \textit{change parameters} (i.e., (m,t)). 
Its main characteristic is the triples that would be added/deleted from the RDF representation of the dataset when such a change occurs. These correspond to the triples that are directly associated with said change and are assumed to be captured by the simple change. 
For example, the above change would be associated with the triple 
\triple{m}{\prange}{t} (which indicates that t is the type of measure m); if said triple is found in \vnew, but not in \vold, then this indicates that a new type (t) has been attached to measure m (thus, Attach\_Type\_To\_Measure(m,t) should be detected).

Triples of this type are required to be in one version but not in the other (i.e., in low-level delta) and are directly associated (\textit{consumed}) by the corresponding simple change. Consumption in this respect means that the corresponding low-level change is captured (described) by said simple change.

A simple change can also have a number of logical conditions required for detection. For example, the triple 
\triple{m}{\prange}{t} may also indicate that a datatype (t) is attached to a dimension m. To determine whether the inclusion of such a triple in \vnew\ corresponds to an 
Attach\_Type\_To\_Measure(m,t) change (as opposed to an
Attach\_Datatype\_To\_Dimension(m,t) change), we need a condition that will determine whether m is a measure or a dimension. In this case, the 
Attach\_Type\_To\_Measure(m,t)
would require the presence of the triple
\triple{m}{\type}{qb:measureProperty} in \vnew.
Note that, in general, conditions may refer to both the old and the new version.
Unlike consumed triples, the triples appearing in conditions are not necessary added or deleted triples; their presence is necessary in either (or both) of the versions and their role is to disambiguate between similar changes.

More formally, a simple change is defined as follows:

\begin{definition}
A \textit{simple change} $c(p_1,\dots,p_n)$ is defined as a tuple of the form $\langle \deltap, \deltam, \cond_{old}, \cond_{new} \rangle$ where:
\begin{itemize}
	\item $c$ is the \textit{name} and $p_1,\dots,p_n \in \vars$, $n \geq 0$, are the \textit{parameters} of the change,
	\item $\deltap, \deltam \subseteq \tps$ are called the \textit{consumed added} and \textit{consumed deleted} triples, respectively, and are sets of triple patterns,
	\item $\cond_{old}, \cond_{new}$ are graph patterns, called the \textit{conditions} related to \vold, \vnew, respectively.
\end{itemize}
\label{def:simple}
\end{definition}

In our running example, Attach\_Type\_To\_Measure(m,t) has two parameters (m, t) and 
$\deltap = \{\triple{m}{\prange}{t}\}$, $\deltam = \emptyset$, $\cond_{old} =$ ``\ '', $\cond_{new} =$ ``\triple{m}{\type}{qb:measureProperty}''.

The structure of Definition~\ref{def:simple} is used for defining the changes that the language of changes accepts, but any actual detection will give specific values (URIs or literals) to the parameters m, t of the change (e.g., Attach\_Type\_To\_Measure(dm-measure:meas7v8t,dm-type:int)). This is captured with the following notion:

\begin{definition}
Consider a change $c(p_1,\dots,p_n)$. Then, an assignment $x_1,\dots, x_n$ of URIs/literals to variables $p_1,\dots,p_n$ is called an \textit{instantiation} of $c$ and denoted by $c(x_1,\dots,x_n)$.
\label{def:simple_instance}
\end{definition}

Now we are in position to formally define the detectability of a change instantiation. 
Intuitively, a change instantiation corresponds to a certain assignment of (some of) the variables in \deltap, \deltam, $\cond_{old}$, $\cond_{new}$; the assignment should be such that the conditions ($\cond_{old}, \cond_{new}$) are true in the underlying datasets, and the triples that correspond to the triple patterns in \deltap, \deltam\ are found in the low-level delta. More precisely:

\begin{definition}
A change instantiation $c(x_1,\dots,x_n)$ of a simple change $c(p_1,\dots,p_n)$ is \textit{detectable} for the pair $\vold, \vnew$ iff 
there is a $\mu \in [[\cond_{old}]]^{\vold} \cap [[\cond_{new}]]^{\vnew}$
such that 
for all $tp \in \deltap$: $\mu(tp) \in \vnew \setminus \vold$ and 
for all $tp \in \deltam$: $\mu(tp) \in \vold \setminus \vnew$ and 
for all $i$: $\mu(p_i) = x_i$. 
\label{def:detectability_simple}
\end{definition}

Then, a detectable change \textit{consumes} the triples that correspond to the triple patterns in \deltap, \deltam. Formally:

\begin{definition}
A detectable change instantiation $c(x_1,\dots,x_n)$ of a simple change $c(p_1,\dots,p_n)$ \textit{consumes} $t \in \vnew \setminus \vold$ (respectively, $t \in \vold \setminus \vnew$) iff there is a $\mu \in [[\cond_{old}]]^{\vold} \cap [[\cond_{new}]]^{\vnew}$
and a $tp \in \deltap$ (respectively, $tp \in \deltam$)
such that $\mu(tp) = t$ and 
for all $i$: $\mu(p_i) = x_i$.
\label{def:consumption_simple}
\end{definition}

The concept of consumption represents the fact that low-level changes are ``assigned'' to simple ones, essentially allowing a grouping (partitioning) of low-level changes into simple ones. To fulfil its purpose, this ``partitioning'' should be perfect, in the sense that all low-level changes should be associated to one, and only one, simple change. This is captured by the properties of \textit{completeness} and \textit{unambiguity}.
Formally:

\begin{definition}
Consider a set of simple changes $C$. 
This set is called \textit{complete} iff 
for any pair of versions \vold, \vnew\ and 
for all $t \in (\vnew \setminus \vold) \cup (\vold \setminus \vnew)$, 
there is an instantiation $c(x_1,\dots,x_n)$ of some $c\in C$ such that $c(x_1,\dots,x_n)$ is detectable and consumes $t$.
\label{def:complete}
\end{definition}

\begin{definition}
Consider a set of simple changes $C$. 
This set is called \textit{unambiguous} iff 
for any pair of versions \vold, \vnew\ and 
for all $t \in (\vnew \setminus \vold) \cup (\vold \setminus \vnew)$,
if $c, c' \in C$ and $c(x_1,\dots,x_n), c'(x_1',\dots,x_m')$ are detectable and consume $t$, then
$c(x_1,\dots,x_n) = c'(x_1',\dots,x_m')$.
\label{def:unambiguous}
\end{definition}

\begin{figure}
	\includegraphics[width=0.49\textwidth]{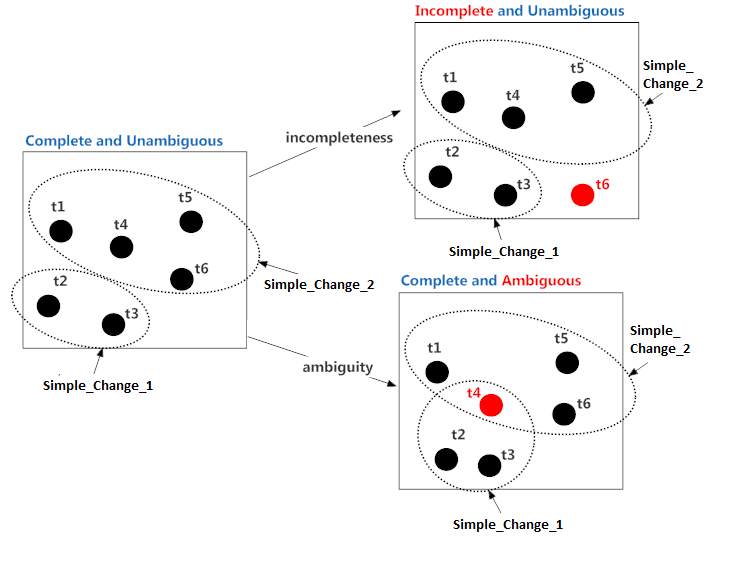}
	\vspace{-0.2in}
	\caption{Visualization of Completeness and Unambiguity}	
	\label{fig:complete}
\end{figure}

In a nutshell, completeness guarantees that all low level changes are associated with at least one simple change, thereby making the reported delta complete (i.e., not missing any change); unambiguity guarantees that no race conditions will emerge between simple changes attempting to consume the same low level change (see Figure~\ref{fig:complete} for a visualization of the notions of completeness and unambiguity). The combination of these two properties guarantees that the delta is produced in a deterministic manner and that it will properly reflect the changes that were actually performed.

\subsection{Complex Changes}
\label{subsec:complexdef}

To guarantee completeness and unambiguity, simple changes must be defined at design time, and remain immutable between different applications; for changes corresponding to custom, application-specific evolution primitives, we use \textit{complex changes}. Complex changes can also be used to define changes that were not foreseen at design-time. This would give an extra flexibility to the user to describe more specialized, coarse-grained changes, which are relevant to a certain application, but not generic enough to be considered part of the ``core'' language of (simple) changes. This essentially allows extending the language of changes at run-time. 

Complex changes are defined in a way similar to simple ones; there are certain differences though.
First, complex changes are not directly associated with low level changes; instead of \deltap, \deltam, they include a set of simple changes, denoted by \deltas, which corresponds to the set of simple changes that should be detectable in order for said complex change to be detectable. This is necessary to make their definition more user-friendly, given that complex changes will be defined by the end-user who does not understand low level changes. 
Second, as complex changes can be freely defined by the user, it would be unrealistic to assume that they will have any quality guarantees, such as completeness or unambiguity. As a consequence, the detection process may lead to non-deterministic consumption of simple changes and conflicts; to avoid this, complex changes are associated with a \textit{priority level}, which is used to resolve such conflicts.

Furthermore, complex changes support \textit{associations}, i.e., correspondences between URIs/literals that are necessary to support changes like 
renames and merge/splits~\cite{DBLP:journals/tods/PapavasileiouFFKC13}.
In general, an association \assoc\ is a pair $(X,Y)$ where $X,Y$ are sets of URIs, literals or variables (called \textit{URI/literal/variable association}, respectively). Associations can have one of the following forms (where $v_i$ may be a URI, literal or variable):
\begin{itemize}
\item $\{v_1\} \rightsquigarrow \{v_2\}$, $v_1 \neq v_2$ (rename)
\item $\{v_0\} \rightsquigarrow \{v_1,\dots,v_n\}$, 
$v_i \neq v_j$ for $i \neq j$ (split)
\item $\{v_1,\dots,v_n\} \rightsquigarrow \{v_0\}$, $v_i \neq v_j$ for $i \neq j$ (merge)
\end{itemize}

A set of URI and literal associations should be provided as an input to the detection process. Such associations could be provided manually by the user, or automatically identified using, e.g., entity matching software~\cite{ksteftut14}; detecting associations is out of the scope of this paper.
Given a mapping $\mu$ and a variable association \assoc, we will denote by $\mu(\assoc)$ the URI/literal association resulting from replacing all variables in \assoc\ with their corresponding URI/literal according to $\mu$.

Complex changes are useful in several applications.
For example, various ontological applications refrain from deleting classes, but use a special subsumption relationship to a class ``Obsolete'' to indicate that a certain class (say, cl) should no longer be used. This action could be captured by the change Mark\_as\_Obsolete(cl), which would consume the simple change Add\_SuperClass(cl, obs), where $obs = geneontology:ObsoleteClass$.

The formal concepts associated with complex changes are similar to the ones related to simple changes:

\begin{definition}
A \textit{complex change} $c(p_1,\dots,p_n)$ is defined as a tuple of the form $\langle \deltas, \cond_{old}, \cond_{new}, \assocset, P \rangle$ where:
\begin{itemize}
	\item $c$ is the \textit{name} and $p_1,\dots,p_n \in \vars$, $n \geq 0$, are the \textit{parameters} of the change,
	\item $\deltas$ is a set of simple changes called the \textit{related simple changes},
	\item $\cond_{old}, \cond_{new}$ are graph patterns, called the \textit{conditions} related to \vold, \vnew, respectively,
	\item $\assocset$ is a set of variable associations,
	\item $P$ is a number corresponding to the \textit{priority level} of the complex change.
\end{itemize}
\label{def:complex}
\end{definition}

In our running example, Mark\_as\_Obsolete(cl) has one parameter (cl) and 
$\deltas = \{ Add\_SuperClass(cl, obs) \}$, $\cond_{old} =$ ``obs = geneontology:ObsoleteClass'', $\cond_{new} =$ ``\ '', $\assocset = \emptyset$, $P=2$.

The definition of complex change instantiations is identical to Definition~\ref{def:simple_instance}. For detectability, we have a series of definitions, so as to take into account priorities:

\begin{definition}
A change instantiation $c(x_1,\dots,x_n)$ of a complex change $c(p_1,\dots,p_n)$ is
\textit{initially detectable} for the pair $\vold$, $\vnew$ and the associations $\assocset^{\vold,\vnew}$ iff 
there is a $\mu \in [[\cond_{old}]]^{\vold} \cap [[\cond_{new}]]^{\vnew}$
such that 
for all $c'(\mu(p_1),$ $\dots,\mu(p_n)) \in \deltas$, $c'(\mu(p_1),$ $\dots,\mu(p_n))$ is detectable,
for all $\assoc \in \assocset$, $\mu(\assoc) \in \assocset^{\vold,\vnew}$ and
for all $i$, $\mu(p_i) = x_i$.
\label{def:init_detectability_complex}
\end{definition}

\begin{definition}
An initially detectable change instantiation $c(x_1,$ $\dots,x_n)$ of a complex change $c(p_1,\dots,p_n)$ \textit{consumes} 
a simple change instantiation $c'(x_1',\dots,x_m')$ of a simple change $c'(p_1',\dots,$ $p_m')$ iff 
$c'(p_1',\dots,p_m') \in \deltas$ and 
there is a $\mu \in [[\cond_{old}]]^{\vold} \cap [[\cond_{new}]]^{\vnew}$ such that
for all $\assoc \in \assocset$, $\mu(\assoc) \in \assocset^{\vold,\vnew}$ and
for all $i$, $\mu(p_i) = x_i, \mu(p_i')=x_i'$.
\label{def:consumption_complex}
\end{definition}

\begin{definition}
A change instantiation $c(x_1,\dots,x_n)$ of a complex change $c(p_1,\dots,p_n)$ is
\textit{detectable} for the pair $\vold,$ $\vnew$ and the associations $\assocset^{\vold,\vnew}$ iff 
it is initially detectable for the pair $\vold,\vnew$ and the associations $\assocset^{\vold,\vnew}$ and
there is no initially detectable change instantiation of another complex change with a higher priority that consumes the same simple change.
\label{def:detectability_complex}
\end{definition}

\section{Representing Changes} 
\label{sec:represent} 

\subsection{Motivation for the Ontology of Changes}

We treat changes as first-class citizens in order to be able to perform queries analysing the evolution of datasets. Further, we are interested in performing combined queries, in which both the datasets and the changes should be considered to get an answer. To achieve this, the representation of the changes that are detected on the data cannot be separated from the data itself. 

For example, consider the following query: ``return all countries for which the unemployment rate of their capital city increased at a rate higher than the average increase of the country as a whole between versions \vold\ and \vnew''. 
This query requires access to the data (to identify countries and capitals) \textit{and} to the changes (to describe the actual increase in the rates of unemployment for each city and country). 
Therefore, to answer it, the changes should be stored in a structured form and their representation should include connections with the actual entities (cities or capitals) that they refer to. 
Note that the definition of a cross-snapshot query language that treats changes as first-class citizens is outside the scope of this paper, however this work provides a formalism on which such a query language can be based. Therefore, we propose representing changes as special entities in an RDF dataset, with connections to the actual data, so that a detectable change can be associated with the corresponding data entities that it refers to.

More specifically, we propose an \textit{ontology of changes} for storing the detected changes, thereby allowing a supervisory look of the detected changes and their association with the entities they refer to in the actual datasets, facilitating the formulation and the answering of queries that refer to both the data and their evolution.

In said ontology, the schema describes the definition of the changes ($c(p_1,\dots,p_n)$ -- see Definitions~\ref{def:simple},~\ref{def:complex}), whereas information on detected changes (which are change instantiations -- see Definition~\ref{def:simple_instance}) appear at the instance level, instantiating the corresponding classes. Specifically, at schema level, we introduce one class for each simple and complex change $c(p_1,\dots,p_n)$ that is understood and considered by the language of changes (\lang), and at instance level, we introduce one individual for each detectable change $c(x_1,\dots,x_n)$ in each pair of versions.

\subsection{Ontology of Changes: Schema Level}

\begin{figure}
\includegraphics[width=0.4\textwidth]{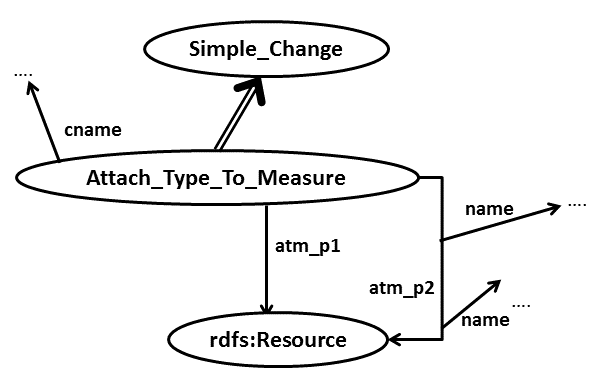}
\caption{Representation of Simple Changes}
\label{fig:simplchan}
\end{figure}

For simple changes, the corresponding change definition is modelled as a subclass of the main class $Simple\_Change$, and is associated with adequate properties to represent its parameters (see Figure~\ref{fig:simplchan}); each such property models the type of the parameter (e.g., whether it is a URI or a literal, via the classes rdfs:Resource, rdfs:Literal respectively) and its name (which is a descriptive name that allows a more intuitive interaction with the user, useful also during the construction of complex changes that consume said simple change). Also a descriptive name ($cname$) is captured for each type of change. Note that conditions do not need to be stored.

For complex changes, similar ideas are used; however, note that the information related to complex changes is generated on the fly at change creation time (in contrast to simple changes, which are in-built in the ontology at design time).
In addition, more detailed information related to the change should be available in the ontology for the change detection process, because, unlike simple changes, this information is not known at design time and cannot be embedded in the code.

\begin{figure}
\includegraphics[width=0.47\textwidth]{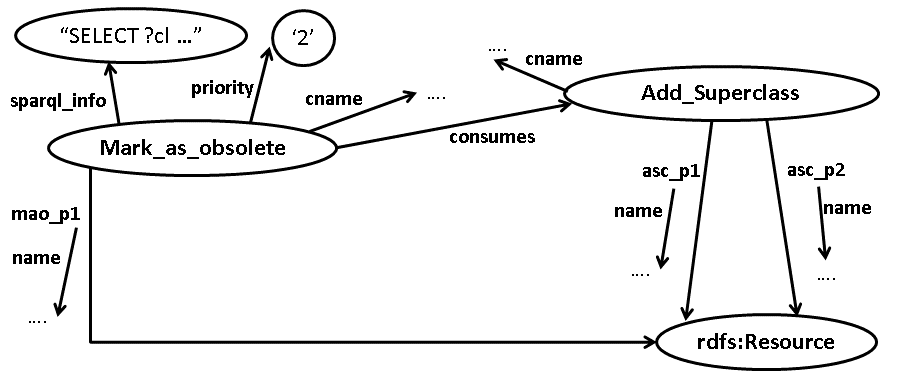}
\caption{Representation of Complex Changes}
\label{fig:comchan}
\end{figure}

Each complex change is modelled as a subclass of the main class $Complex\_Change$, and is associated with adequate properties to represent its parameters (see Figure~\ref{fig:comchan}).
Again, parameters have a type and a name, which are modelled in the same manner as in simple changes.
In addition, each complex change is associated with the simple change(s) that it consumes, and includes also properties describing its (user-defined) descriptive name and its priority.

Finally, for each complex change, the SPARQL query used for its detection, is automatically generated at change definition time; this is done for efficiency, to avoid having to generate this query in every run of the detection process. 
The SPARQL query encapsulates all the relevant information about the detection of the change, so further information on the complex change (namely conditions and associations) need not to be stored.

\subsection{Ontology of Changes: Data Level}

\begin{figure}
\includegraphics[width=83mm]{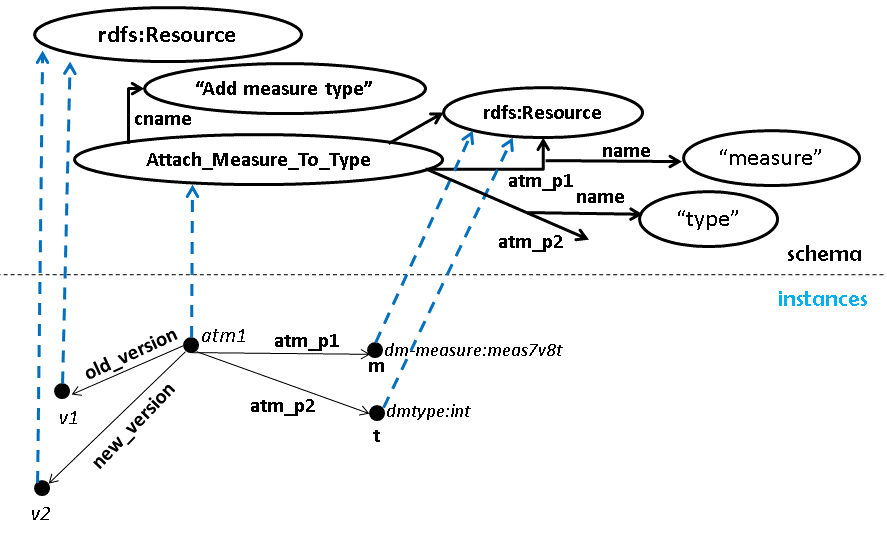}
\caption{Representation of Simple Change Detection}
\label{fig:sc detection}
\end{figure}

\begin{figure}
\includegraphics[width=83mm]{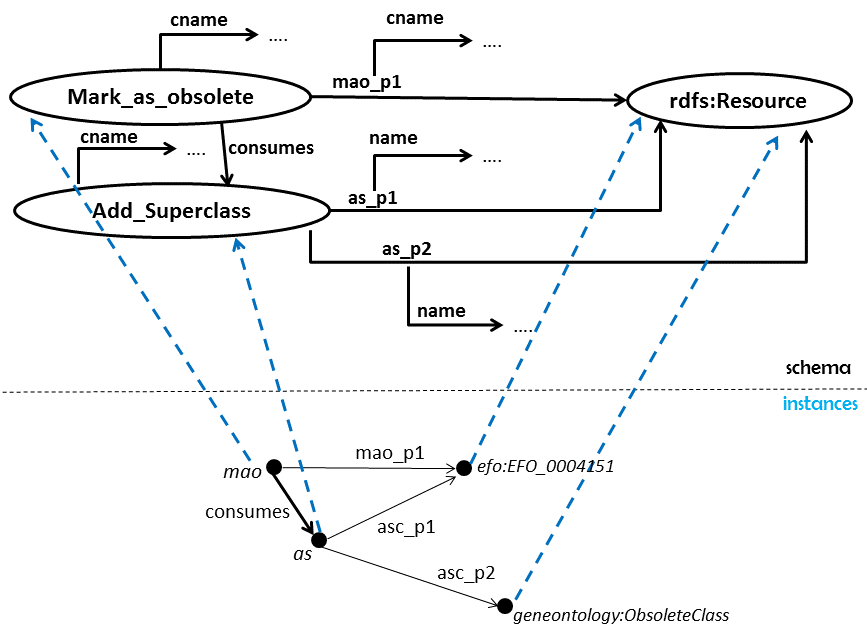}
\caption{Representation of Complex Change Detection}
\label{fig:cc detection}
\end{figure}

The instance level is used to store the detectable simple and complex changes.
Each detectable change between any two versions is represented using a different individual associated through adequate properties with all relevant information, namely, the versions between which the change was detected and the exact values of its parameters (of the change instantiation). 
For complex changes, we additionally need to include consumption information.
Examples of such instantiations for simple and complex changes are shown in Figures~\ref{fig:sc detection} and~\ref{fig:cc detection}, for the detectable changes Attach\_Type\_To\_Mea- sure($dm-measure:meas7v8t,dm-type:int$) and \newline
Mark\_as\_Obsolete ($efo:EFO\_0004151$), respectively.

\subsection{Ontology of Changes: Associations}

\begin{figure}
\includegraphics[width=0.46\textwidth]{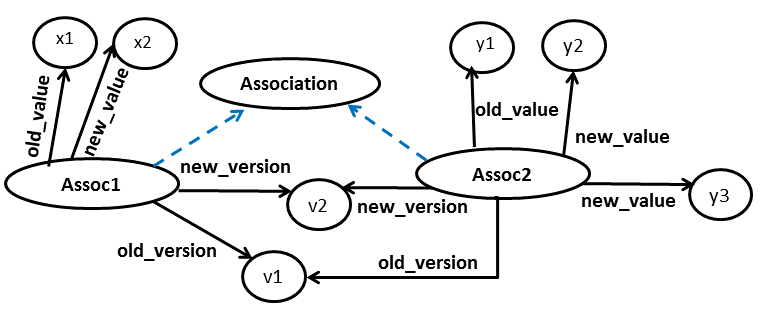}
\caption{Representation of Associations}
\label{fig:assoc_inst}
\end{figure}

Another useful feature of the ontology of changes is the storage of \emph{associations}, which are necessary for the detection of complex changes.
Associations are stored as instances classified under the class $Association$, and record the versions between which they are applicable, as well as the old and new values in the corresponding association. For example, Figure~\ref{fig:assoc_inst} shows the representation of the associations $\{x1\} \rightsquigarrow \{x2\}$ and $\{y1\} \rightsquigarrow \{y2, y3\}$, which appear between versions $v1,v2$.

\section{Detecting and Storing Changes} 
\label{sec:detect} 

The change detection process is responsible for the detection of simple and complex changes between two dataset versions (and their corresponding associations, which are a necessary input for the detection of complex changes), as well as for the enrichment of the ontology of changes with information about the detectable changes.
Thus, the process can be considered to comprise of two steps: the first is the identification of the detectable changes and the corresponding information (triples) to be inserted in the ontology of changes (\textit{triple creation}), and the second is the actual ingestion of said information in the ontology of changes (\textit{triple ingestion}). 

To detect simple and complex changes, we rely on plain SPARQL queries, which are generated from the information drawn from the definition of the corresponding changes 
(Definitions~\ref{def:detectability_simple},~\ref{def:detectability_complex}).
For simple changes, this information is known at design time, so the query is loaded from a configuration file, whereas for complex changes, the corresponding query is generated once at change-creation time (run-time) and is loaded from the ontology of changes (see Figure~\ref{fig:comchan}).
The results of the generated queries determine the change instantiations that are detectable; this information is further processed to determine the actual triples to be inserted in the ontology of changes. Recall that this depends on the actual detected change and its type (see Section~\ref{sec:represent}).

More specifically, the SPARQL queries used for detecting a simple change are \texttt{SELECT} queries, whose returned values are the parameter values of the change; thus, for each parameter of the change definition, we put one variable in the \texttt{SELECT} clause. Then, the \texttt{WHERE} clause of the query includes the triple patterns that should (or should not) be found in each of the versions in order for a change instantiation to be detectable; more specifically, the triple patterns in \deltap\ must be found in \vnew\ but not in \vold, the triple patterns in \deltam\ must be found in \vold\ but not in \vnew, and the graph patterns in $\cond_{old},\cond_{new}$ should be applied in \vold, \vnew, respectively.

Let's use this methodology to construct the SPARQL query for the simple change 
Attach\_Type\_To\_Measure(m,t) in which: 
$\deltap = \{\triple{m}{\prange}{t}\}$, $\deltam = \emptyset$, 
$\cond_{old} =$ ``\ '', $\cond_{new} =$ ``\triple{m}{\type}{qb:measureProperty}''.
The corresponding SPARQL query is:\vspace{0.1in}\\
\texttt{ 
SELECT ?m ?t WHERE \{\\ 
GRAPH \vnew  \{ ?m rdf:type qb:MeasureProperty. \\
?m rdfs:range ?t. \} \\ 
FILTER NOT EXISTS \\
\{ GRAPH \vold  \{ ?m rdfs:range ?t. \} \} \\
\} \\
}

Each result row of this query corresponds to the parameter values of one detectable change instantiation; for example, if the query returns the pair (dm-measure:meas7v8t,dm-type:int), then the change instantiation
Attach\_Type\_To\_Measure(dm-measure:meas7v8t,dm-type:int) is detectable.

The generation of the SPARQL queries for the complex changes follows a similar pattern. One difference is that the \texttt{WHERE} clause should also include the required associations, which are assumed to be already stored in the ontology of changes.
A second important difference is that complex changes check the existence of simple changes (namely, those found in the \deltas\ part of the complex change definition) in the ontology of changes, rather than triples in the two versions (as is the case with simple changes detection); therefore, complex changes should be detected after the detection of simple changes and their storage in the ontology.
Note also that the considered simple changes should not have been marked as ``consumed'' by other detectable changes of a higher priority; thus, it is important for queries associated with complex changes to be executed in a particular order, as implied by their priority.

Let's illustrate the above methodology to construct the SPARQL query for the complex change 
Mark\_as\_Obsolete(cl). This change has one parameter (cl) and 
$\deltas = \{ Add\_SuperClass(cl, obs) \}$, 
$\cond_{old} =$ ``obs = geneontology:ObsoleteClass'', $\cond_{new} =$ ``\ '', $\assocset = \emptyset$, $P=2$. 
The corresponding SPARQL query is: \vspace{0.1in}\\
\texttt{
SELECT ?cl WHERE \{ \\
GRAPH <changesOntology> \{ \\ 
?asc a co:Add\_Superclass; \\
	co:asc\_p1 ?cl; \\
	co:asc\_p2 ?obs. \\
FILTER NOT EXISTS \{ ?cc co:consumes ?asc \} \\
FILTER (?obs = ``geneontology:ObsoleteClass'').  \\
\} \\
\}
}

As with simple changes, each query result corresponds to the parameter values of a detectable change instantiation; for example, if the query finds the simple change 
Add\_Superclass(\url{efo:EFO_0004151}, \url{geneontology:ObsoleteClass}) which is not consumed by any complex change, 
then it will return 
\url{efo:EFO_0004151}, 
which implies that 
the change instantiation \newline
Mark \_as\_Obsolete(\url{efo:EFO_0004151}) is detectable.

Following detection, the information about the detectable (simple or complex) change instantiations must be stored in the ontology of changes. To do so, 
we have to process each result row to create the corresponding triple blocks, as specified in 
Section~\ref{sec:represent}. 
This is done as a separate process that first stores the triple blocks in a file (on disk) and subsequently uploads them in Virtuoso using Virtuoso' bulk loading process (triple ingestion).

Note that the detection and storing of changes could be done in one step, if one used an adequately defined SPARQL update statement that identified the detectable change instantiations, created the corresponding triple blocks and inserted them in the ontology using a single statement. 
However, this approach turned out to be slower by 1-2 orders of magnitude, partly because it does not exploit bulk updates based on multiple threads, and also because bulk loading is much faster than inserting triples using SPARQL updates in Virtuoso.

The described process enjoys the characteristics put forward in the previous sections, namely:
\begin{description}
\item [Expressiveness:] The algorithm supports the entire semantics of SPARQL, thereby allowing a very expressive method to define changes.
\item [Extensibility:] The addition of a new change (simple or complex) requires just the addition of the corresponding SPARQL query, with no further modifications on the source code.
\item [Data Model Insensitivity:] The process can be applied in any data model (e.g., multi-dimensional) which can be expressed in RDF (see Section~\ref{sec:apply}) ignoring its internal representation; all peculiarities of the underlying data model are incorporated in the corresponding SPARQL queries, thereby allowing the code to be versatile.
\item [Cross-Platform Character:] The process requires very generic RDF data management operations (SPARQL support and RDF data import), so it can be implemented in any triple store. 
\end{description}

\section{Application of the Framework}
\label{sec:apply}


As mentioned above, our framework is flexible and generic enough to be applicable to any data model, which is transformable to the RDF format.
Thus, to apply our framework to any given data model, one first has to determine how to transform said data model in RDF. 
The second step is the definition of the simple changes that would comprise the \textit{language of changes} (\lang); note that said changes should be complete and unambiguous for optimal results. 
Defining the simple changes amounts to identifying the SPARQL query that is necessary for detection, as well as the representation of the simple changes in the schema of the changes ontology; the latter can be easily automated.

Complex changes need not be defined at this point, as they are created at run-time; to facilitate the definition of complex changes one could create an adequate user-friendly interface, that could (potentially) sacrifice some of the expressive power of the complex change definition (Definition~\ref{def:complex}) in favour of user-friendliness; at any rate, the definition of such an interface is beyond the scope of this work.


Indicatively, we demonstrate, for visualization and experimentation purposes, the above process for the RDF and multi-dimensional data models;
a similar methodology could be used for other data models, e.g., relational, but further applications of our framework are omitted due to space limitations.

\subsection{RDF Model}

The RDF data model is a popular model for exposing, sharing, and connecting pieces of data, information, and knowledge on the Semantic Web. Scientists from various areas of expertise use RDF to publish their scientific observations and measurements on the Web. 

The case of RDF is straightforward in the sense that no data transformation is required.
For the definition of simple changes, we used a set very similar to the so-called ``basic changes'' in~\cite{DBLP:journals/tods/PapavasileiouFFKC13}.
The full list of defined simple changes, along with details on their definition and the corresponding SPARQL queries used for detection can be found
\ExtOrSub
{in Appendix \ref{ext:rdfsc}.}
{in the extended version of this paper~\cite{extV}.}


\subsection{Multi-dimensional Model}
The multi-dimensional data model is useful for capturing statistics and information on data items along several dimensions.
The bridge to RDF is achieved through the Data Cube vocabulary\footnote{http://www.w3.org/TR/vocab-data-cube} which was proposed as a W3C recommendation to enable the publication of  statistical data flows and other multi-dimensional data sets over the Web. 
The Data Cube vocabulary allows multi-dimensional data to enjoy all the advantages of the LOD and RDF technologies (knowledge sharing and interconnection) via its publication to RDF, thereby allowing publishers or third parties to annotate and link to specified data, which can be flexibly combined across different datasets.

To apply our framework to the multi-dimensional model, we adopt the transformation proposed by the Data Cube vocabulary. 
The definition of simple changes is based on the identification of all entities of the Data Cube vocabulary (i.e., fact tables, dimensions, observations, codelists, hierarchies, measures, attributes) in order to express special or generic cases of changes that appear in each of those entities.
To achieve completeness and unambiguity in our definition of simple changes, we take into account 
how data cube entities are connected and which sets of triple patterns 
denote the existence/characteristics of the corresponding entities.
Again, the full list of the defined simple changes for the multi-dimensional model, along with details on their definition and the corresponding SPARQL queries used for detection can be found
\ExtOrSub
{in Appendix \ref{ext:mdsc}.}
{in the extended version of this paper~\cite{extV}.}


\section{Experimental Evaluation}
\label{sec:eval}

\begin{table}[t!]
\small
\centering
\caption{Evaluated Datasets: Versions and Sizes}
\begin{tabular}{|c|c|c|}
\hline 
\textbf{Dataset}& 
\textbf{Version}& 
\textbf{\# Triples} \\ \hline \hline

\multicolumn{3}{|c|}{\textbf{RDF datasets}} \\ \hline


\multirow{3}{*}{ATLAS} & v12.07 & 457.951.940 \\
& v13.05 & 422.144.126 \\
& v13.07 & 447.149.655 \\ 
\hline

\multirow{3}{*}{Dbpedia} & v3.7 & 48.898.490 \\
& v3.8 & 63.126.304 \\
& v3.9 & 67.980.265 \\ 
\hline

\multirow{3}{*}{GO} & 24-03-2009 & 189.378 \\
& 22-09-2009 & 195.125 \\
& 20-04-2010 & 210.076 \\ 
\hline

\multicolumn{3}{|c|}{\textbf{Multi-dimensional datasets}} \\ \hline

\multirow{3}{*}{4lqc} & v1 & 229.338.925 \\
& v2 & 243.336.594 \\
& v3 & 243.449.874 \\
\hline

\multirow{3}{*}{1zph} & v1 & 4.022.424 \\
& v2 & 4.022.352 \\
& v3 & 4.022.208 \\ 
\hline

\multirow{3}{*}{1bu4} & v1 & 259.125 \\
& v2 & 270.049 \\
& v3 & 270.049 \\
\hline

\end{tabular}
\label{tbl:info_datasets}
\end{table}

Our framework was implemented and applied on the data models described in Section~\ref{sec:apply}. The evaluation considered some representative real-world datasets of various sizes from these two data models, and its aims were the following:
\begin{itemize}
\item To identify the number and type of simple changes that usually occur in real world settings.
\item To study the performance of our change detection process when dealing with real settings and quantify the effect of the size of the compared versions and the number of detected changes in the performance of the algorithm.
\end{itemize}
To our knowledge, this is the first time that change detection has been evaluated for datasets of this size and versatility (RDF and multi-dimensional). 

\begin{figure*}
\includegraphics[width=\textwidth]{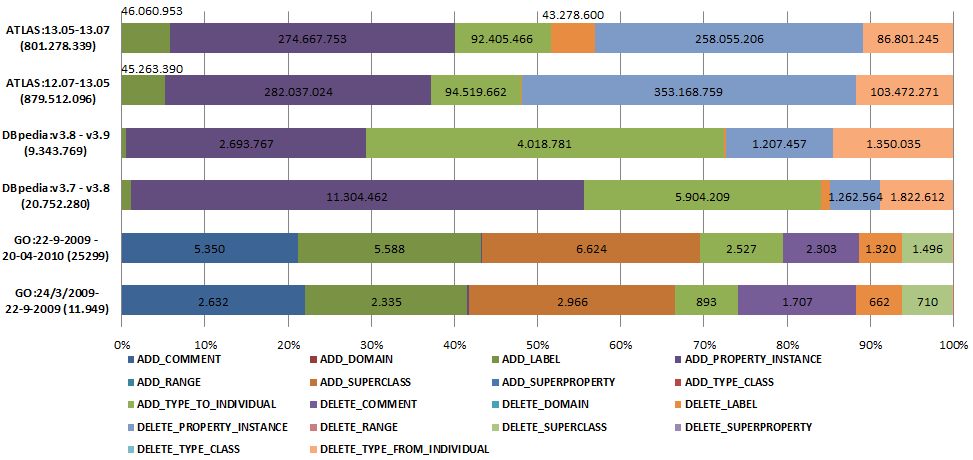}
\includegraphics[width=\textwidth]{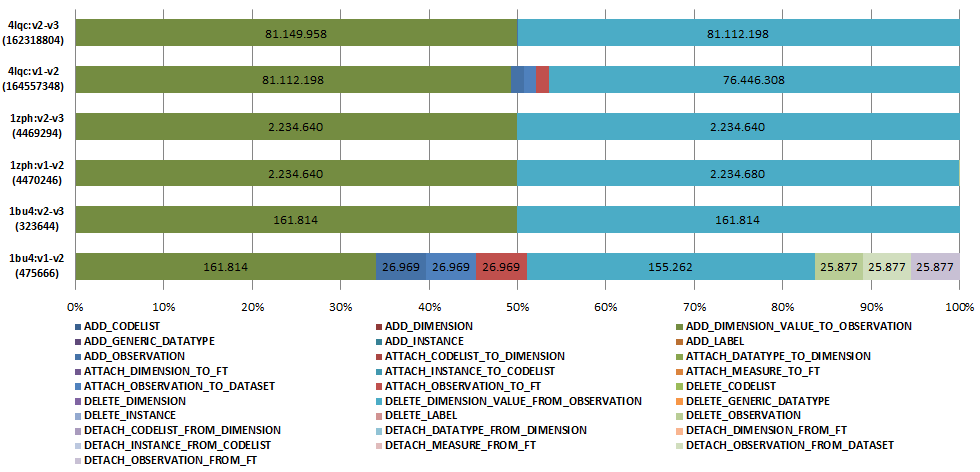}
\caption{Detected Changes (RDF/Multi-dimensional)}
\label{fig:results_changes}
\end{figure*}

\begin{table*}
\small
\centering
\caption{Performance of Change Detection (RDF/Multi-dimensional)}
\begin{tabular}{|c|c|c|c|c|c|}
\hline
\textbf{Datasets and Versions} &
\textbf{\# Simple Changes} & 
\textbf{\# Ingested Triples} & 
\textbf{Triple Creation (sec)} &
\textbf{Triple Ingestion (sec)} &
\textbf{Duration (sec)} \\	\hline \hline

\multicolumn{6}{|c|}{\textbf{RDF datasets}} \\ \hline


ATLAS: v12.07-v13.05
& 879.512.096	& 4.980.846.857		& 7.294,73	& 9.654,46	& 16.949,20	\\ \hline

ATLAS: v13.05-v13.07
& 801.278.339	& 4.539.113.055		& 5.953,07	& 9.243,61	& 15.196,68	\\ \hline

DBpedia: v3.7-v3.8
& 20.752.280		& 116.327.560          & 465,26	& 143,49	& 608,74	\\ \hline

DBpedia: v3.8-v3.9	
& 9.343.769		& 50.620.418				& 296,20	& 72,94		& 369,14	\\ \hline

GO: 24-03-2009-20-04-2010 
& 11.949	& 59.179	& 0,80	& 0,50	& 1,31	\\ \hline

GO: 22-09-2009-20-04-2010
& 25.299	& 124.262	& 0,96	& 0,57	& 1,52	\\ \hline

\multicolumn{6}{|c|}{\textbf{Multi-dimensional datasets}} \\ \hline

4lqc: v1-v2
& 164.557.348	& 978.012.302	& 1.651,03	& 994,85	& 2.645,87	\\ \hline

4lqc: v2-v3
& 162.318.804	& 973.837.296	& 1.645,03	& 996,96	& 2.641,99	\\ \hline

1zph: v1-v2
& 4.469.294		& 26.815.814		& 129,38		& 46,91		& 176,29	\\ \hline

1zph: v2-v3
& 4.469.262		& 26.815.494		& 131,25		& 47,33		& 178,58	\\ \hline

1bu4: v1-v2
& 475.666			& 2.642.549		& 7,42			& 6,60		& 14,02 	\\ \hline

1bu4: v2-v3	
& 323.644			& 1.941.848		& 5,12			& 5,97		& 11,09		\\ \hline

\end{tabular}
\label{tbl:res_perf}
\end{table*}

\subsection{Setting}
\label{subsec:evalset}

For the management of linked data (e.g., storage of datasets and query execution), we worked with a scalable triple store, namely the open source version of 
Virtuoso Universal Server\footnote{\url{http://virtuoso.openlinksw.com}}, v7.10.3209 (note that, our work is not bounded to any specific infrastructure or triple-store). 
Virtuoso is hosted on a machine which uses an Intel Xeon E5-2630 at 2.30GHz, with 384GB of RAM running Debian Linux wheeze version, with Linux kernel 3.16.4. The system uses 7TB RAID-5 HDD configurations. 
From the total amount of memory, we dedicated 64GB for Virtuoso and 5GB for the implemented application. 
Moreover, taking into account that CPU provides 12 cores with 2 threads each, we decided to use a multi-threaded implementation; specifically, we noticed that the use of 8 threads during the creation of the RDF triples along with the ingestion process gave us optimal results for our setting. This was one more reason to select Virtuoso for our implementation, as it allows the concurrent use of multiple threads during ingestion. 
To eliminate the effects of hot/cold starts, cached OS information etc., each change detection process was executed 10 times and the average times were considered.

For our experimental evaluation, we used 3 RDF and 3 multi-dimensional datasets. 
The selected RDF datasets were 
ATLAS\footnote{\url{http://www.ebi.ac.uk/gxa/home}}, 
a subset of the English 
Dbpedia\footnote{\url{http://dbpedia.org}} 
(consisting of article categories, instance types, labels and mapping-based properties) 
and
GO\footnote{\url{http://geneontology.org}}.
The selected multi-dimensional datasets were provided by 
Datamarket and contain various statistics; in particular, 
4lqc\footnote{\url{https://datamarket.com/data/set/4lqc}} contains weather measurements from the Global Historical Climatology Network, 
1buh\footnote{\url{https://datamarket.com/data/set/1buh}} contains numbers of employed persons between 15 and 64 years old taking time off over the last 12 months for family sickness or emergencies and 1zph\footnote{\url{https://datamarket.com/data/set/1zph}} contains information about unemployment by sex, age, duration of unemployment and registration. 
All these measurements are taken from Eurostat. 
Table~\ref{tbl:info_datasets} summarizes the corresponding versions and their sizes.

\subsection{Detected Changes}

The first part of our evaluation studies the number and type of simple changes that appear in the evaluated datasets.
The results are summarized in Figure~\ref{fig:results_changes}, where we note 
the large number of changes which occurred during ATLAS evolution compared to the other datasets. 
This is explained by the fact that ATLAS contains experimental biological results and measurements that change over time, thus new versions are vastly different from previous ones.
Moreover, note that the majority of changes (in all datasets except GO) are applied to the data level (e.g., 
Delete\_Property\_Instance), whereas in GO, we have also changes which are applied to the schema.

On the multi-dimensional model, the majority of changes in all datasets are of type Add\_Dimension\_Value\_to\_Observation and De- lete\_Dimension\_Value\_from\_Observation. For both types of chan- ges, we observe that almost the same number of changes is reported, which implies that the datasets' evolution mainly consists of changing dimension values upon observations; thus, creating a complex change to capture this pair of changes would immediately halve the number of reported changes.

\subsection{Performance of Change Detection} 
\label{subsec:chdet_perf}

The second part of our evaluation examines the performance of the detection process for the RDF and multi-dimensional testbeds; the results appear in Table~2. 
The reported performance is split into two parts, namely triple creation and triple ingestion; as explained in Section~\ref{sec:detect}, the former includes the execution of the SPARQL queries for detection and the identification of the triples to be inserted in the ontology of changes, whereas the latter is the actual enrichment of the ontology of changes.

The main conclusion from Table~2 is that 
the number of simple changes is a more crucial factor for performance than the sizes of the compared versions (see also Table~\ref{tbl:info_datasets}). 
This observation is more clear in the DBpedia dataset, where the evolution between v3.7 and v3.8 produces about twice the number of changes than the evolution between v3.8 and v3.9; despite the fact that in the second case, we have to compare larger dataset versions, the execution time in the former case is almost twice as large.
Note that this conclusion holds for both triple creation and ingestion.


More importantly, the performance of our approach is about 1 order of magnitude faster than the performance reported by~\cite{DBLP:journals/tods/PapavasileiouFFKC13}, upon which this work was built, even if the reported simple changes are compatible in both works, as the used simple changes for the RDF model are similar to the so-called ``basic changes'' in~\cite{DBLP:journals/tods/PapavasileiouFFKC13}. 
For example, when comparing the performance of change detection over the GO versions v22-09-2009 and v20-04-2010, our approach needs 1,52 sec, while \cite{DBLP:journals/tods/PapavasileiouFFKC13} requires 33,13 sec.

\section{Related Work}
\label{sec:related}


In general, approaches for change detection can be classified into low-level and high-level ones, based on the types of changes they support. 
Low-level change detection approaches report simple add/delete operations, which are not concise or intuitive enough to human users, while focusing on machine readability. \cite{franconi10} discusses a low-level detection approach for propositional Knowledge Bases (KBs), which can be easily extended to apply to KBs represented under any classical knowledge representation formalism. This work presents a number of desirable formal properties for change detection languages, like delta uniqueness, reversibility of changes and the ability to move backwards and forwards in the history of versions using the deltas. 
Similar properties appear in~\cite{Zeginis:2011:CDR:1993053.1993056}, where a low-level change detection formalism for RDFS datasets is presented, as well as in~\cite{DBLP:journals/tods/PapavasileiouFFKC13}, upon which this work builds. 

\cite{Konev:2008:LDP:1431108.1431137} describes a low-level change detection approach for the Description Logic $\mathcal{E}L$; there, the focus is on a concept-based description of changes, and the returned delta is a set of concepts whose position in the class hierarchy changed. \cite{Kontchakov08} presents a formal low-level change detection approach for DL-Lite ontologies, which focuses on a semantical description of the changes. Recently, \cite{journals/jis/ImLK13} introduced a scalable approach for reasoning-aware low-level change detection that uses a relational database management system, while~\cite{Dong2007} supports change detection between RDF datasets containing blank nodes. All these works result in non-concise, low-level deltas, which are difficult for a human to understand.

High-level change detection approaches provide more human-readable deltas. Although there is no agreed-upon list of changes that are necessary for any given context, various high-level operations, along with the intuition behind them, have been proposed~\cite{klein04,Noy:2002:PFA:777092.777207,Plessers:2007:UOE:1229184.1229198}. 
However, these approaches do not present formal semantics of such operations, or of the corresponding detection process; thus, no useful formal properties can be guaranteed. 

In \cite{klein04,Noy:2002:PFA:777092.777207}, a fixed-point algorithm for detecting changes, implemented in PromptDiff, is described. The algorithm incorporates heuristic-based matchers to detect changes between two versions, thus introducing uncertainty in the results, and obtaining a recall of 96\% and a precision of 93\%. 

\cite{Plessers:2007:UOE:1229184.1229198} proposes the Change Definition Language (CDL) as a means to define high-level changes. A change is defined and detected using temporal queries over a version log that contains recordings of the applied low-level changes. The version log must be updated whenever a change occurs; this overrules the use of this approach in non-curated or distributed environments. In our work, version logs are not necessary for the detection, as the delta can be produced a posteriori. \cite{Auer:2006:VEF:1760700.1760710} focuses on defining a formal way to represent high-level changes as sequences of triples, but does not describe a detection process or a specific language of changes.
Finally, \cite{Staab} proposes an interesting high-level change detection algorithm that takes into account the semantics of OWL.

\section{Conclusions}
\label{sec:conclusions}

In this paper, we proposed an approach to cope with the dynamicity of Web datasets via the management of changes between versions. We advocated in favour of a flexible, extendible and triple-store independent approach, which is suitable for any data model that is representable in RDF terms. Our approach prescribes the definition of data model-specific, as well as application-specific, changes, and their management (definition, storage, detection) in a manner that ensures the satisfaction of formal properties (like completeness and unambiguity), the flexibility and customization of the considered changes (via complex changes, which can be defined at run-time), as well as the easy configuration of a scalable detection mechanism (via a generic algorithm that builds on  SPARQL queries that can be easily generated from the changes' definitions).

This work builds upon our previous work on the topic of change detection~\cite{DBLP:journals/tods/PapavasileiouFFKC13}, by providing a more generic change definition framework, that is significantly more scalable and versatile than the one proposed in~\cite{DBLP:journals/tods/PapavasileiouFFKC13}, as shown in Section~\ref{sec:eval}.

\ExtOrSub{
\section{Acknowledgments}
This work was partially supported by the EU FP7 project DIACHRON (ICT-2011.4.3, \#601043).
The authors would like to thank G. Briem for providing the multi-dimensional datasets necessary for our experiments, as well as M. Meimaris, T. Galani and C. Pateritsas for support and discussions in previous versions of this work.
}
{}

\clearpage

\bibliographystyle{abbrv} 
\bibliography{WWW15} 

\ExtOrSub{
\clearpage
\appendix
\section{Change Detection Analysis}
\label{app:changes}

\subsection{Ontological Data Changes Analysis}
The following tables provide the change detection analysis for all the considered datasets. 

\begin{table}[h]
\caption{DBpedia Dataset}
\begin{tabular}{r|c|c|}
\cline{2-3}
\multicolumn{1}{l|}{}                                         & \multicolumn{1}{l|}{\textbf{\begin{tabular}[c]{@{}l@{}}v3.7-\\ v3.8\end{tabular}}} & 
\multicolumn{1}{l|}{\textbf{\begin{tabular}[c]{@{}l@{}}v3.8-\\ v3.9\end{tabular}}} \\ \hline
\multicolumn{1}{|r|}{\textbf{ADD\_COMMENT}}                   & 0                                        & 0                                        \\ \hline
\multicolumn{1}{|r|}{\textbf{ADD\_DOMAIN}}                    & 0                                        & 0                                         \\ \hline
\multicolumn{1}{|r|}{\textbf{ADD\_LABEL}}                     & 229805                                      & 49399                                       \\ \hline
\multicolumn{1}{|r|}{\textbf{\begin{tabular}[c]{@{}r@{}}ADD\_PROPERTY\_\\ INSTANCE\end{tabular}}}        & 11304462                                     & 2693767                                     \\ \hline
\multicolumn{1}{|r|}{\textbf{ADD\_RANGE}}                     & 0                                         & 0                                        \\ \hline
\multicolumn{1}{|r|}{\textbf{ADD\_SUPERCLASS}}                & 0                                       & 0                                       \\ \hline
\multicolumn{1}{|r|}{\textbf{ADD\_SUPERPROPERTY}}             & 0                                           & 0                                           \\ \hline
\multicolumn{1}{|r|}{\textbf{ADD\_TYPE\_CLASS}}               & 0                                        & 0                                        \\ \hline
\multicolumn{1}{|r|}{\textbf{\begin{tabular}[c]{@{}r@{}}ADD\_TYPE\_TO\_\\ INDIVIDUAL\end{tabular}}}      & 5904209                                      & 4018781                                      \\ \hline
\multicolumn{1}{|r|}{\textbf{DELETE\_COMMENT}}                & 0                                      & 0                                        \\ \hline
\multicolumn{1}{|r|}{\textbf{DELETE\_DOMAIN}}                 & 0                                         & 0                                       \\ \hline
\multicolumn{1}{|r|}{\textbf{DELETE\_LABEL}}                  & 228628                                       & 24330                                       \\ \hline
\multicolumn{1}{|r|}{\textbf{\begin{tabular}[c]{@{}r@{}}DELETE\_PROPERTY\_\\ INSTANCE\end{tabular}}}     & 1262564                                     & 1207457                                    \\ \hline
\multicolumn{1}{|r|}{\textbf{DELETE\_RANGE}}                  & 0                                         & 0 \\ \hline
\multicolumn{1}{|r|}{\textbf{DELETE\_SUPERCLASS}}             & 0                                     & 0                                       \\ \hline
\multicolumn{1}{|r|}{\textbf{DELETE\_SUPERPROPERTY}}          & 0                                           & 0                                           \\ \hline
\multicolumn{1}{|r|}{\textbf{DELETE\_TYPE\_CLASS}}            & 0                                     & 0                                       \\ \hline
\multicolumn{1}{|r|}{\textbf{\begin{tabular}[c]{@{}r@{}}DELETE\_TYPE\_\\ FROM\_INDIVIDUAL\end{tabular}}} & 1822612                                      & 1350035                                      \\ \hline
\end{tabular}
\end{table}

\begin{table}[h]
\caption{GO Dataset}
\begin{tabular}{r|c|c|}
\cline{2-3}
\multicolumn{1}{l|}{}                                         & \multicolumn{1}{l|}{\textbf{\begin{tabular}[c]{@{}l@{}}go:24/3/2009-\\ 22-9-2009\end{tabular}}} & 
\multicolumn{1}{l|}{\textbf{\begin{tabular}[c]{@{}l@{}}go:22-9-2009-\\ 20-04-2010\end{tabular}}} \\ \hline
\multicolumn{1}{|r|}{\textbf{ADD\_COMMENT}}                   & 2632                            & 5350                             \\ \hline
\multicolumn{1}{|r|}{\textbf{ADD\_DOMAIN}}                    & 0                               & 0                                \\ \hline
\multicolumn{1}{|r|}{\textbf{ADD\_LABEL}}                     & 2335                            & 5588                             \\ \hline
\multicolumn{1}{|r|}{\textbf{\begin{tabular}[c]{@{}r@{}}ADD\_PROPERTY\_\\ INSTANCE\end{tabular}}}        & 22                              & 36                               \\ \hline
\multicolumn{1}{|r|}{\textbf{ADD\_RANGE}}                     & 0                               & 0                                \\ \hline
\multicolumn{1}{|r|}{\textbf{ADD\_SUPERCLASS}}                & 2966                            & 6624                             \\ \hline
\multicolumn{1}{|r|}{\textbf{ADD\_SUPERPROPERTY}}             & 0                               & 0                                \\ \hline
\multicolumn{1}{|r|}{\textbf{ADD\_TYPE\_CLASS}}               & 0                               & 0                                \\ \hline
\multicolumn{1}{|r|}{\textbf{\begin{tabular}[c]{@{}r@{}}ADD\_TYPE\_TO\_\\ INDIVIDUAL\end{tabular}}}      & 893                             & 2527                             \\ \hline
\multicolumn{1}{|r|}{\textbf{DELETE\_COMMENT}}                & 1707                            & 2303                             \\ \hline
\multicolumn{1}{|r|}{\textbf{DELETE\_DOMAIN}}                 & 0                               & 0                                \\ \hline
\multicolumn{1}{|r|}{\textbf{DELETE\_LABEL}}                  & 662                             & 1320                             \\ \hline
\multicolumn{1}{|r|}{\textbf{\begin{tabular}[c]{@{}r@{}}DELETE\_PROPERTY\_\\ INSTANCE\end{tabular}}}    & 0                               & 0                                \\ \hline
\multicolumn{1}{|r|}{\textbf{DELETE\_RANGE}}                  & 0                               & 0                                \\ \hline
\multicolumn{1}{|r|}{\textbf{DELETE\_SUPERCLASS}}             & 710                             & 1496                             \\ \hline
\multicolumn{1}{|r|}{\textbf{DELETE\_SUPERPROPERTY}}          & 0                               & 0                                \\ \hline
\multicolumn{1}{|r|}{\textbf{DELETE\_TYPE\_CLASS}}            & 0                               & 0                                \\ \hline
\multicolumn{1}{|r|}{\textbf{\begin{tabular}[c]{@{}r@{}}DELETE\_TYPE\_\\ FROM\_INDIVIDUAL\end{tabular}}} & 22                              & 55                               \\ \hline
\end{tabular}
\end{table}

\begin{table}[h]
\caption{Atlas Dataset}
\begin{tabular}{r|c|c|}
\cline{2-3}
\multicolumn{1}{l|}{}                                                                                    & \multicolumn{1}{l|}{\textbf{\begin{tabular}[c]{@{}l@{}}atlas:12.07-\\ 13.05\end{tabular}}} 
& \multicolumn{1}{l|}{\textbf{\begin{tabular}[c]{@{}l@{}}atlas:13.05-\\ 13.07\end{tabular}}} \\ \hline
\multicolumn{1}{|r|}{\textbf{ADD\_COMMENT}}                                                              & 707                                                                                        & 79                                              \\ \hline
\multicolumn{1}{|r|}{\textbf{ADD\_DOMAIN}}                                                               & 42                                                                                         & 18                                              \\ \hline
\multicolumn{1}{|r|}{\textbf{ADD\_LABEL}}                                                                & 45263390                                                                                   & 46060953                                        \\ \hline
\multicolumn{1}{|r|}{\textbf{\begin{tabular}[c]{@{}r@{}}ADD\_PROPERTY\_\\ INSTANCE\end{tabular}}}        & 282037024                                                                                  & 274667753                                       \\ \hline
\multicolumn{1}{|r|}{\textbf{ADD\_RANGE}}                                                                & 48                                                                                         & 20                                              \\ \hline
\multicolumn{1}{|r|}{\textbf{ADD\_SUPERCLASS}}                                                           & 18455                                                                                      & 5814                                            \\ \hline
\multicolumn{1}{|r|}{\textbf{ADD\_SUPERPROPERTY}}                                                        & 239                                                                                        & 18                                              \\ \hline
\multicolumn{1}{|r|}{\textbf{ADD\_TYPE\_CLASS}}                                                          & 12574                                                                                      & 1813                                            \\ \hline
\multicolumn{1}{|r|}{\textbf{\begin{tabular}[c]{@{}r@{}}ADD\_TYPE\_TO\_\\ INDIVIDUAL\end{tabular}}}      & 94519662                                                                                   & 92405466                                        \\ \hline
\multicolumn{1}{|r|}{\textbf{DELETE\_COMMENT}}                                                           & 32                                                                                         & 79                                              \\ \hline
\multicolumn{1}{|r|}{\textbf{DELETE\_DOMAIN}}                                                            & 0                                                                                          & 6                                               \\ \hline
\multicolumn{1}{|r|}{\textbf{DELETE\_LABEL}}                                                             & 1017308                                                                                    & 43278600                                        \\ \hline
\multicolumn{1}{|r|}{\textbf{\begin{tabular}[c]{@{}r@{}}DELETE\_PROPERTY\_\\ INSTANCE\end{tabular}}}     & 353168759                                                                                  & 258055206                                       \\ \hline
\multicolumn{1}{|r|}{\textbf{DELETE\_RANGE}}                                                             & 0                                                                                          & 6                                               \\ \hline
\multicolumn{1}{|r|}{\textbf{DELETE\_SUPERCLASS}}                                                        & 1003                                                                                       & 1112                                            \\ \hline
\multicolumn{1}{|r|}{\textbf{DELETE\_SUPERPROPERTY}}                                                     & 14                                                                                         & 16                                              \\ \hline
\multicolumn{1}{|r|}{\textbf{DELETE\_TYPE\_CLASS}}                                                       & 568                                                                                        & 135                                             \\ \hline
\multicolumn{1}{|r|}{\textbf{\begin{tabular}[c]{@{}r@{}}DELETE\_TYPE\_\\ FROM\_INDIVIDUAL\end{tabular}}} & 103472271                                                                                  & 86801245                                        \\ \hline
\end{tabular}
\end{table}

\subsection{Multidimensional Data Changes Analysis}
The following tables provide the change detection analysis for all the considered datasets. 

\begin{table}[h]
\caption{1bu4 Datasets}
\begin{tabular}{r|c|c|}
\cline{2-3}
\textbf{}                                                                                                             & \multicolumn{1}{l|}{\textbf{v1-v2}} & \multicolumn{1}{l|}{\textbf{v2-v3}} \\ \hline
\multicolumn{1}{|r|}{\textbf{ADD\_CODELIST}}                                                                          & 1                                   & 0                                   \\ \hline
\multicolumn{1}{|r|}{\textbf{ADD\_DIMENSION}}                                                                         & 1                                   & 0                                   \\ \hline
\multicolumn{1}{|r|}{\textbf{\begin{tabular}[c]{@{}r@{}}ADD\_DIMENSION\_VALUE\_\\ TO\_OBSERVATION\end{tabular}}}      & 161814                              & 161814                              \\ \hline
\multicolumn{1}{|r|}{\textbf{ADD\_GENERIC\_DATATYPE}}                                                                 & 1                                   & 0                                   \\ \hline
\multicolumn{1}{|r|}{\textbf{ADD\_INSTANCE}}                                                                          & 6                                   & 0                                   \\ \hline
\multicolumn{1}{|r|}{\textbf{ADD\_LABEL}}                                                                             & 2                                   & 0                                   \\ \hline
\multicolumn{1}{|r|}{\textbf{ADD\_OBSERVATION}}                                                                       & 26969                               & 0                                   \\ \hline
\multicolumn{1}{|r|}{\textbf{ATTACH\_CODELIST\_TO\_DIMENSION}}                                                        & 1                                   & 0                                   \\ \hline
\multicolumn{1}{|r|}{\textbf{ATTACH\_DATATYPE\_TO\_DIMENSION}}                                                        & 1                                   & 0                                   \\ \hline
\multicolumn{1}{|r|}{\textbf{ATTACH\_DIMENSION\_TO\_FT}}                                                              & 7                                   & 7                                   \\ \hline
\multicolumn{1}{|r|}{\textbf{ATTACH\_INSTANCE\_TO\_CODELIST}}                                                         & 6                                   & 0                                   \\ \hline
\multicolumn{1}{|r|}{\textbf{ATTACH\_MEASURE\_TO\_FT}}                                                                & 1                                   & 1                                   \\ \hline
\multicolumn{1}{|r|}{\textbf{\begin{tabular}[c]{@{}r@{}}ATTACH\_OBSERVATION\_\\ TO\_DATASET\end{tabular}}}            & 26969                               & 0                                   \\ \hline
\multicolumn{1}{|r|}{\textbf{ATTACH\_OBSERVATION\_TO\_FT}}                                                            & 26969                               & 0                                   \\ \hline
\multicolumn{1}{|r|}{\textbf{DELETE\_CODELIST}}                                                                       & 1                                   & 0                                   \\ \hline
\multicolumn{1}{|r|}{\textbf{DELETE\_DIMENSION}}                                                                      & 1                                   & 0                                   \\ \hline
\multicolumn{1}{|r|}{\textbf{\begin{tabular}[c]{@{}r@{}}DELETE\_DIMENSION\_VALUE\_\\ FROM\_OBSERVATION\end{tabular}}} & 155262                              & 161814                              \\ \hline
\multicolumn{1}{|r|}{\textbf{DELETE\_GENERIC\_DATATYPE}}                                                              & 1                                   & 0                                   \\ \hline
\multicolumn{1}{|r|}{\textbf{DELETE\_INSTANCE}}                                                                       & 5                                   & 0                                   \\ \hline
\multicolumn{1}{|r|}{\textbf{DELETE\_LABEL}}                                                                          & 2                                   & 0                                   \\ \hline
\multicolumn{1}{|r|}{\textbf{DELETE\_OBSERVATION}}                                                                    & 25877                               & 0                                   \\ \hline
\multicolumn{1}{|r|}{\textbf{\begin{tabular}[c]{@{}r@{}}DETACH\_CODELIST\_FROM\_\\ DIMENSION\end{tabular}}}           & 1                                   & 0                                   \\ \hline
\multicolumn{1}{|r|}{\textbf{\begin{tabular}[c]{@{}r@{}}DETACH\_DATATYPE\_FROM\_\\ DIMENSION\end{tabular}}}           & 1                                   & 0                                   \\ \hline
\multicolumn{1}{|r|}{\textbf{DETACH\_DIMENSION\_FROM\_FT}}                                                            & 7                                   & 7                                   \\ \hline
\multicolumn{1}{|r|}{\textbf{\begin{tabular}[c]{@{}r@{}}DETACH\_INSTANCE\_FROM\_\\ CODELIST\end{tabular}}}            & 5                                   & 0                                   \\ \hline
\multicolumn{1}{|r|}{\textbf{DETACH\_MEASURE\_FROM\_FT}}                                                              & 1                                   & 1                                   \\ \hline
\multicolumn{1}{|r|}{\textbf{\begin{tabular}[c]{@{}r@{}}DETACH\_OBSERVATION\_FROM\_\\ DATASET\end{tabular}}}          & 25877                               & 0                                   \\ \hline
\multicolumn{1}{|r|}{\textbf{\begin{tabular}[c]{@{}r@{}}DETACH\_OBSERVATION\_\\ FROM\_FT\end{tabular}}}               & 25877                               & 0                                   \\ \hline
\end{tabular}
\end{table}

\begin{table}[h]
\caption{1zph Dataset}
\begin{tabular}{r|c|c|}
\cline{2-3}
\multicolumn{1}{l|}{}                                                                                                 & \textbf{v1 - v2} & \textbf{v2 - v3} \\ \hline
\multicolumn{1}{|r|}{\textbf{\begin{tabular}[c]{@{}r@{}}ADD\_DIMENSION\_VALUE\_\\ TO\_OBSERVATION\end{tabular}}}      & 2234640          & 2234640          \\ \hline
\multicolumn{1}{|r|}{\textbf{ADD\_OBSERVATION}}                                                                       & 148              & 0                \\ \hline
\multicolumn{1}{|r|}{\textbf{ATTACH\_DIMENSION\_TO\_FT}}                                                              & 6                & 6                \\ \hline
\multicolumn{1}{|r|}{\textbf{ATTACH\_MEASURE\_TO\_FT}}                                                                & 1                & 1                \\ \hline
\multicolumn{1}{|r|}{\textbf{\begin{tabular}[c]{@{}r@{}}ATTACH\_OBSERVATION\_TO\_\\ DATASET\end{tabular}}}            & 148              & 0                \\ \hline
\multicolumn{1}{|r|}{\textbf{ATTACH\_OBSERVATION\_TO\_FT}}                                                            & 148              & 0                \\ \hline
\multicolumn{1}{|r|}{\textbf{\begin{tabular}[c]{@{}r@{}}DELETE\_DIMENSION\_VALUE\_\\ FROM\_OBSERVATION\end{tabular}}} & 2234680          & 2234640          \\ \hline
\multicolumn{1}{|r|}{\textbf{DELETE\_OBSERVATION}}                                                                    & 156              & 0                \\ \hline
\multicolumn{1}{|r|}{\textbf{DETACH\_DIMENSION\_FROM\_FT}}                                                            & 6                & 6                \\ \hline
\multicolumn{1}{|r|}{\textbf{DETACH\_MEASURE\_FROM\_FT}}                                                              & 1                & 1                \\ \hline
\multicolumn{1}{|r|}{\textbf{\begin{tabular}[c]{@{}r@{}}DETACH\_OBSERVATION\_FROM\_\\ DATASET\end{tabular}}}          & 156              & 0                \\ \hline
\multicolumn{1}{|r|}{\textbf{DETACH\_OBSERVATION\_FROM\_FT}}                                                          & 156              & 0                \\ \hline
\end{tabular}
\end{table}

\begin{table}[h]
\caption{4lqc Dataset}
\begin{tabular}{r|c|c|}
\cline{2-3}
\multicolumn{1}{l|}{}                                                                                                 & \textbf{v1 - v2} & \textbf{v2 - v3} \\ \hline
\multicolumn{1}{|r|}{\textbf{\begin{tabular}[c]{@{}r@{}}ADD\_DIMENSION\_VALUE\_\\ TO\_OBSERVATION\end{tabular}}}      & 81112198         & 81149958         \\ \hline
\multicolumn{1}{|r|}{\textbf{ADD\_OBSERVATION}}                                                                       & 2332944          & 18880            \\ \hline
\multicolumn{1}{|r|}{\textbf{ATTACH\_DIMENSION\_TO\_FT}}                                                              & 3                & 3                \\ \hline
\multicolumn{1}{|r|}{\textbf{ATTACH\_MEASURE\_TO\_FT}}                                                                & 1                & 1                \\ \hline
\multicolumn{1}{|r|}{\textbf{\begin{tabular}[c]{@{}r@{}}ATTACH\_OBSERVATION\_\\ TO\_DATASET\end{tabular}}}            & 2332945          & 18880            \\ \hline
\multicolumn{1}{|r|}{\textbf{ATTACH\_OBSERVATION\_TO\_FT}}                                                            & 2332945          & 18880            \\ \hline
\multicolumn{1}{|r|}{\textbf{\begin{tabular}[c]{@{}r@{}}DELETE\_DIMENSION\_VALUE\_\\ FROM\_OBSERVATION\end{tabular}}} & 76446308         & 81112198         \\ \hline
\multicolumn{1}{|r|}{\textbf{DETACH\_DIMENSION\_FROM\_FT}}                                                            & 3                & 3                \\ \hline
\multicolumn{1}{|r|}{\textbf{DETACH\_MEASURE\_FROM\_FT}}                                                              & 1                & 1                \\ \hline
\end{tabular}
\end{table}
\clearpage
\section{Simple Changes for RDF data model}
\label{ext:rdfsc}

In this appendix, we list the simple changes referring to the RDF model. For each change and taking into account Definition~\ref{def:simple} we present: 
\begin{itemize}
	\item Its name.
	\item The intuition it captures, described in terms of the RDF model. 
	\item Its parameters, and the intuition behind each parameter.
	\item The SPARQL query which will be used for its detection.
	\item The consumed added and deleted triples $\deltap, \deltam \subseteq \tps$ which are essentially sets of triple patterns.	
	\item The conditions related to \vold, \vnew which form the respective graph patterns $\cond_{old}, \cond_{new}$.
\end{itemize}
For simplicity, we will write v1 to denote \vold and v2 to denote \vnew in the following tables. Moreover, the conditions use clauses like ``\uu{a} does not appear in v1'', rather than the more verbose (and formal) graph pattern: 
$\cond_{old}$=``FILTER NOT EXISTS \{ \uu{a} ?p1 ?o1. ?s2 ?p2 \uu{a}. ?s3 \uu{a} ?o3\}".

\begin{table*}[h!]
\begin{footnotesize}
\begin{tabular}{|p{2.0cm}|p{5.5cm}|p{5.5cm}|}
\hline
Change &
\textbf{\op{Add\_Type\_Class($a$)}} &
\textbf{\op{Delete\_Type\_Class($a$)}} \\ \hline

Intuition &
Add object $a$ of type \class &
Delete object $a$ of type \class \\ \hline

Parameters &
$a$ = The added object &
$a$ = The deleted object\\ \hline

SPARQL used for detection &
\vspace{-0.4cm}
\begin{verbatim}
SELECT ?a WHERE { 
GRAPH <v2> { 
?a rdf:type rdf:Class. }
FILTER NOT EXISTS { 
GRAPH <v1> {
?a rdf:type rdf:Class. }
} }
\end{verbatim}
\vspace{-0.4cm}
&
\vspace{-0.4cm}
\begin{verbatim}
SELECT ?a WHERE { 
GRAPH <v1> { 
?a rdf:type rdf:Class.}
FILTER NOT EXISTS { 
GRAPH <v2> {
?a rdf:type rdf:Class.}
} }
\end{verbatim}
\vspace{-0.4cm}
\\ \hline

$\chng^+$ 
&
\triple{a}{\type}{\class}
&
$\emptyset$ \\ \hline

$\chng^-$ &
$\emptyset$  &
\triple{a}{\type}{\class} \\ \hline

$\cond_{old}$ &
\uu{a} does not appear in v1 &
--\\ \hline

$\cond_{new}$ &
-- &
\uu{a} does not appear in v2 \\ \hline

\end{tabular}
\end{footnotesize}
\newline
\end{table*}

\begin{table*}[h]
\begin{footnotesize}
\begin{tabular}{|p{2.0cm}|p{5.5cm}|p{5.5cm}|}
\hline
Change &
\textbf{\op{Add\_Type\_Property($a$)}} &
\textbf{\op{Delete\_Type\_Property($a$)}} \\ \hline

Intuition &
Add object $a$ of type \prop &
Delete object $a$ of type \prop \\ \hline

Parameters &
$a$ = The added object &
$a$ = The deleted object  \\ \hline

SPARQL used for detection &
\vspace{-0.4cm}
\begin{verbatim}
SELECT ?a WHERE { 
GRAPH <v2> { 
?a rdf:type rdf:Property. }
FILTER NOT EXISTS { 
GRAPH <v1> {
?a rdf:type rdf:Property. }
} }
\end{verbatim}
\vspace{-0.4cm}
&
\vspace{-0.4cm}
\begin{verbatim}
SELECT ?a WHERE { 
GRAPH <v1> { 
?a rdf:type rdf:Property.}
FILTER NOT EXISTS { 
GRAPH <v2> {
?a rdf:type rdf:Property.}
} }
\end{verbatim}
\vspace{-0.4cm}
\\ \hline

$\chng^+$ 
&
\triple{a}{\type}{\prop}
&
$\emptyset$ \\ \hline

$\chng^-$ &
$\emptyset$  &
\triple{a}{\type}{\prop} \\ \hline

$\cond_{old}$ &
\uu{a} does not appear in v1 &
--\\ \hline

$\cond_{new}$ &
-- &
\uu{a} does not appear in v2 \\ \hline

\end{tabular}
\end{footnotesize}
\newline
\newline
The changes \op{Add\_Type\_Individual} and
\op{Delete\_Type\_Individual}
are defined analogously with the exception that \newline \triple{a}{\type}{\res} should be in $\chng^+$ ($\chng^-$) instead of \triple{a}{\type}{\prop}.
\end{table*}

\begin{table*}[h]
\begin{footnotesize}
\begin{tabular}{|p{2.0cm}|p{5.5cm}|p{5.5cm}|}
\hline
Change &
\textbf{\op{Add\_Superclass($a$,$b$)}} &
\textbf{\op{Delete\_Superclass($a$,$b$)}}  \\ \hline

Intuition &
Parent $b$ of class $a$ is added &
Parent $b$ of class $a$ is deleted\\ \hline

Parameters &
$a$ = The class \newline
$b$ = The new parent  &
$a$ = The class \newline
$b$ = The old parent \\ \hline

SPARQL used for detection &
\vspace{-0.4cm}
\begin{verbatim}
SELECT ?a ?b WHERE { 
GRAPH <v2> { 
?a rdfs:subClassOf ?b } 
FILTER NOT EXISTS { 
GRAPH <v1> { 
?a rdfs:subClassOf ?b } 
} }
\end{verbatim}
\vspace{-0.4cm}
&
\vspace{-0.4cm}
\begin{verbatim}
SELECT ?a ?b WHERE { 
GRAPH <v1> { 
?a rdfs:subClassOf ?b } 
FILTER NOT EXISTS { 
GRAPH <v2> { 
?a rdfs:subClassOf ?b } 
} }
\end{verbatim}
\vspace{-0.4cm}
\\ \hline

$\chng^+$ 
&
\triple{a}{\cisa}{b}
&
$\emptyset$ \\ \hline

$\chng^-$ &
$\emptyset$  &
\triple{a}{\cisa}{b} \\ \hline

$\cond_{old}$ &
-- &
--\\ \hline

$\cond_{new}$ &
\triple{a}{\type}{\class}. \newline
\triple{b}{\type}{\class}
&
\triple{a}{\type}{\class}. \newline
\triple{b}{\type}{\class}
\\ \hline

\end{tabular}
\end{footnotesize}
\newline
\end{table*}

\begin{table*}[h]
\begin{footnotesize}
\begin{tabular}{|p{2.0cm}|p{5.5cm}|p{5.5cm}|}
\hline
Change &
\textbf{\op{Add\_Superproperty($a$,$b$)}} &
\textbf{\op{Delete\_Superproperty($a$,$b$)}}  \\ \hline

Intuition &
Parent $b$ of property $a$ is added &
Parent $b$ of property $a$ is deleted \\ \hline

Parameters &
$a$ = The property \newline
$b$ = The new parent  &
$a$ = The property \newline
$b$ = The old parent  \\ \hline

SPARQL used for detection &
\vspace{-0.4cm}
\begin{verbatim}
SELECT ?a ?b WHERE { 
GRAPH <v2> { 
?a rdfs:subPropertyOf ?b } 
FILTER NOT EXISTS { 
GRAPH <v1> { 
?a rdfs:subPropertyOf ?b } 
} }
\end{verbatim}
\vspace{-0.4cm}
&
\vspace{-0.4cm}
\begin{verbatim}
SELECT ?a ?b WHERE { 
GRAPH <v1> { 
?a rdfs:subPropertyOf ?b } 
FILTER NOT EXISTS { 
GRAPH <v2> { 
?a rdfs:subPropertyOf ?b } 
} }
\end{verbatim}
\vspace{-0.4cm}
\\ \hline

$\chng^+$ 
&
\triple{a}{\pisa}{b}
&
$\emptyset$ \\ \hline

$\chng^-$ &
$\emptyset$  &
\triple{a}{\pisa}{b} \\ \hline

$\cond_{old}$ &
-- &
-- \\ \hline

$\cond_{new}$ &
\triple{a}{\type}{\prop}.\newline
\triple{b}{\type}{\prop} 
&
\triple{a}{\type}{\prop}.\newline
\triple{b}{\type}{\prop}
\\ \hline

\end{tabular}
\end{footnotesize}
\end{table*}

\begin{table*}[h]
\begin{footnotesize}
\begin{tabular}{|p{2.0cm}|p{5.5cm}|p{5.5cm}|}
\hline
Change &
\textbf{\op{Add\_Type\_To\_Individual($a$,$b$)}} &
\textbf{\op{Delete\_Type\_From\_Individual($a$,$b$)}}  \\ \hline

Intuition &
Type $b$ of individual $a$ is added &
Type $b$ of individual $a$ is deleted \\ \hline

Parameters &
$a$ = The individual \newline
$b$ = The new type (class) &
$a$ = The individual \newline
$b$ = The old type (class) \\ \hline

SPARQL used for detection &
\vspace{-0.4cm}
\begin{verbatim}
SELECT ?a ?b WHERE { 
GRAPH <v2> { ?a rdf:type ?b. 
FILTER (?b != rdf:Class && 
?b != rdfs:Property && 
?b != rdfs:Resource). } 
FILTER NOT EXISTS { 
GRAPH <v1> { ?a rdf:type ?b. 
FILTER (?b != rdf:Class && 
?b != rdfs:Property && 
?b != rdfs:Resource). } } 
} 
\end{verbatim}
\vspace{-0.4cm}
&
\vspace{-0.4cm}
\begin{verbatim}
SELECT ?a ?b WHERE { 
GRAPH <v1> { ?a rdf:type ?b. 
FILTER (?b != rdf:Class && 
?b != rdfs:Property && 
?b != rdfs:Resource). } 
FILTER NOT EXISTS { 
GRAPH <v2> { ?a rdf:type ?b. 
FILTER (?b != rdf:Class && 
?b != rdfs:Property && 
?b != rdfs:Resource). } } 
}  
\end{verbatim}
\vspace{-0.4cm}
\\ \hline

$\chng^+$ 
&
\triple{a}{\type}{b}
&
$\emptyset$ \\ \hline

$\chng^-$ &
$\emptyset$  &
\triple{a}{\type}{b} \\ \hline

$\cond_{old}$ &
-- &
-- \\ \hline

$\cond_{new}$ &
\triple{a}{\type}{\res}. \newline
FILTER (?b != rdf:Class \&\& \newline
?b != rdfs:Property \&\& \newline
?b != rdfs:Resource).
&
\triple{a}{\type}{\res}. \newline
FILTER (?b != rdf:Class \&\& \newline
?b != rdfs:Property \&\& \newline
?b != rdfs:Resource).
\\ \hline

\end{tabular}
\end{footnotesize}
\newline
\end{table*}

\begin{table*}[h]
\begin{footnotesize}
\begin{tabular}{|p{2.0cm}|p{5.5cm}|p{5.5cm}|}
\hline
Change &
\textbf{\op{Add\_Property\_Instance($a_1$,$a_2$,$b$)}} &
\textbf{\op{Delete\_Property\_Instance($a_1$,$a_2$,$b$)}} \\ \hline

Intuition &
Add property instance of property $b$ &
Delete property instance of property $b$ \\ \hline

Parameters &
$a_1$ = The subject \newline
$a_2$ = The object \newline
$b$ = The property &
$a_1$ = The subject \newline
$a_2$ = The object \newline
$b$ = The property \\ \hline

SPARQL used for detection &
\vspace{-0.4cm}
\begin{verbatim}
SELECT ?a1 ?b ?a2 WHERE { 
GRAPH <v2> { ?a1 ?b ?a2. 
FILTER(?b!=rdfs:subClassOf && 
?b!=rdfs:subPropertyOf && 
?b!=rdf:type && 
?b!=rdfs:comment && 
?b!=rdfs:label && 
?b!=rdfs:domain && 
?b!=rdfs:range) } 
FILTER NOT EXISTS { 
GRAPH <v1> { ?a1 ?b ?a2. 
FILTER(?b!=rdfs:subClassOf && 
?b!=rdfs:subPropertyOf && 
?b!=rdf:type && 
?b!=rdfs:comment && 
?b!=rdfs:label && 
?b!=rdfs:domain && 
?b!=rdfs:range) } } 
} 
\end{verbatim}
\vspace{-0.4cm}
&
\vspace{-0.4cm}
\begin{verbatim}
SELECT ?a1 ?b ?a2 WHERE { 
GRAPH <v1> { ?a1 ?b ?a2. 
FILTER(?b!=rdfs:subClassOf && 
?b!=rdfs:subPropertyOf && 
?b!=rdf:type && 
?b!=rdfs:comment && 
?b!=rdfs:label && 
?b!=rdfs:domain && 
?b!=rdfs:range) } 
FILTER NOT EXISTS { 
GRAPH <v2> { ?a1 ?b ?a2. 
FILTER(?b!=rdfs:subClassOf && 
?b!=rdfs:subPropertyOf && 
?b!=rdf:type && 
?b!=rdfs:comment && 
?b!=rdfs:label && 
?b!=rdfs:domain && 
?b!=rdfs:range) } } 
} 
\end{verbatim}
\vspace{-0.4cm}
\\ \hline

$\chng^+$ 
&
\triple{a1}{b}{a2}
&
$\emptyset$ \\ \hline

$\chng^-$ &
$\emptyset$  &
\triple{a1}{b}{a2} \\ \hline

$\cond_{old}$ &
-- &
-- \\ \hline

$\cond_{new}$ &
FILTER (?b != rdfs:subClassOf \&\& \newline
?b != rdfs:subPropertyOf \&\& \newline
?b != rdf:type \&\& ?b != rdfs:comment \&\& \newline
?b != rdfs:label \&\& ?b != rdfs:domain \&\& ?b != rdfs:range)
&
FILTER (?b != rdfs:subClassOf \&\& \newline
?b != rdfs:subPropertyOf \&\& \newline
?b != rdf:type \&\& ?b != rdfs:comment \&\& \newline
?b != rdfs:label \&\& ?b != rdfs:domain \&\& ?b != rdfs:range)
\\ \hline

\end{tabular}
\end{footnotesize}
\end{table*}

\begin{table*}[h]
\begin{footnotesize}
\begin{tabular}{|p{2.0cm}|p{5.5cm}|p{5.5cm}|}
\hline
Change &
\textbf{\op{Add\_Domain($a$,$b$)}} &
\textbf{\op{Delete\_Domain($a$,$b$)}}  \\ \hline

Intuition &
Domain $b$ of property $a$ is added &
Domain $b$ of property $a$ is deleted \\ \hline

Parameters &
$a$ = The property  \newline
$b$ = The domain &
$a$ = The property  \newline
$b$ = The domain \\ \hline

SPARQL used for detection &
\vspace{-0.4cm}
\begin{verbatim}
SELECT ?a ?b WHERE { 
GRAPH <v2> { 
?a rdfs:domain ?b }. 
FILTER NOT EXISTS { 
GRAPH <v1> { 
?a rdfs:domain ?b } } 
}
\end{verbatim}
\vspace{-0.4cm}
&
\vspace{-0.4cm}
\begin{verbatim}
SELECT ?a ?b WHERE { 
GRAPH <v1> { 
?a rdfs:domain ?b }. 
FILTER NOT EXISTS { 
GRAPH <v2> { 
?a rdfs:domain ?b } } 
}
\end{verbatim}
\vspace{-0.4cm}
\\ \hline

$\chng^+$ 
&
\triple{a}{\pdomain}{b}
&
$\emptyset$ \\ \hline

$\chng^-$ &
$\emptyset$  &
\triple{a}{\pdomain}{b} \\ \hline

$\cond_{old}$ &
-- &
-- \\ \hline

$\cond_{new}$ &
\triple{a}{\type}{\prop}
&
\triple{a}{\type}{\prop}
\\ \hline

\end{tabular}
\end{footnotesize}
\newline
\newline
The changes \op{Add\_Range} and
\op{Delete\_Range}
are defined analogously with the exception that \triple{a}{\prange}{b} \newline
should be in $\chng^+$ ($\chng^-$) instead of \triple{a}{\pdomain}{b}.
\end{table*}

\begin{table*}[h]
\begin{footnotesize}
\begin{tabular}{|p{2.0cm}|p{5.5cm}|p{5.5cm}|}
\hline
Change &
\textbf{\op{Add\_Comment($a$,$b$)}} &
\textbf{\op{Delete\_Comment($a$,$b$)}}  \\ \hline

Intuition &
Comment $b$ of object $a$ is added &
Comment $b$ of object $a$ is deleted \\ \hline

Parameters &
$a$ = The object  \newline
$b$ = The new comment &
$a$ = The object  \newline
$b$ = The old comment  \\ \hline

SPARQL used for detection &
\vspace{-0.4cm}
\begin{verbatim}
SELECT ?a ?b WHERE { 
GRAPH <v2> { 
?a rdfs:comment ?b }. 
FILTER NOT EXISTS { 
GRAPH <v1> { 
?a rdfs:comment ?b } } 
}
\end{verbatim}
\vspace{-0.4cm}
&
\vspace{-0.4cm}
\begin{verbatim}
SELECT ?a ?b WHERE { 
GRAPH <v1> { 
?a rdfs:comment ?b }. 
FILTER NOT EXISTS { 
GRAPH <v2> { 
?a rdfs:comment ?b } } 
}
\end{verbatim}
\vspace{-0.4cm}
\\ \hline

$\chng^+$ 
&
\triple{a}{\comment}{b}
&
$\emptyset$ \\ \hline

$\chng^-$ &
$\emptyset$  &
\triple{a}{\comment}{b} \\ \hline

$\cond_{old}$ &
-- &
-- \\ \hline

$\cond_{new}$ &
--
&
--
\\ \hline

\end{tabular}
\end{footnotesize}
\end{table*}

\begin{table*}[h]
\begin{footnotesize}
\begin{tabular}{|p{2.0cm}|p{5.5cm}|p{5.5cm}|}
\hline
Change &
\textbf{\op{Add\_Label($a$,$b$)}} &
\textbf{\op{Delete\_Label($a$,$b$)}}  \\ \hline

Intuition &
Label $b$ of object $a$ is added &
Label $b$ of object  $a$ is deleted \\ \hline

Parameters &
$a$ = The object  \newline
$b$ = The new comment &
$a$ = The object  \newline
$b$ = The old comment  \\ \hline

SPARQL used for detection &
\vspace{-0.4cm}
\begin{verbatim}
SELECT ?a ?b WHERE { 
GRAPH <v2> { 
?a rdfs:label ?b }. 
FILTER NOT EXISTS { 
GRAPH <v1> { 
?a rdfs:label ?b } } 
}
\end{verbatim}
\vspace{-0.4cm}
&
\vspace{-0.4cm}
\begin{verbatim}
SELECT ?a ?b WHERE { 
GRAPH <v1> { 
?a rdfs:label ?b }. 
FILTER NOT EXISTS { 
GRAPH <v2> { 
?a rdfs:label ?b } } 
}
\end{verbatim}
\vspace{-0.4cm}
\\ \hline

$\chng^+$ 
&
\triple{a}{\labell}{b}
&
$\emptyset$ \\ \hline

$\chng^-$ &
$\emptyset$  &
\triple{a}{\labell}{b} \\ \hline

$\cond_{old}$ &
-- &
--\\ \hline

$\cond_{new}$ &
--
&
--
\\ \hline

\end{tabular}
\end{footnotesize}
\end{table*}

\clearpage

\section{Simple Changes for Multidimensional data model}
\label{ext:mdsc}

In this appendix, we list the simple changes refering to the multidimensional model. SPARQL queries have been expressed in RDF Data Cube vocabulary as soon as the data have been transformed respectively to this format.

 

For each change and taking into account Definition~\ref{def:simple} we present: 
\begin{itemize}
	\item Its name.
	\item The intuition it captures, described in terms of the RDF model. 
	\item Its parameters, and the intuition behind each parameter.
	\item The SPARQL query which will be used for its detection.
	\item The consumed added and deleted triples $\deltap, \deltam \subseteq \tps$ which are essentially sets of triple patterns.	
\end{itemize}



\begin{table*}[h]
\begin{footnotesize}
\begin{tabular}{|p{2cm}|p{6.5cm}|p{6.5cm}|}
\hline
Change &
\textbf{\op{Add\_Dimension($d$)}} &
\textbf{\op{Delete\_Dimension($d$)}} \\ \hline

Intuition &
Add a new dimension (not assigned to a fact table) &
Delete a dimension \\ \hline

Parameters &
$d$ = The added dimension &
$d$ = The deleted dimension\\ \hline

SPARQL used for detection &
\vspace{-0.4cm}
\begin{verbatim}
SELECT ?d WHERE { 
GRAPH <v2> { 
?d a qb:DimensionProperty. 
} 
FILTER NOT EXISTS { GRAPH <v1> { 
?d a qb:DimensionProperty. 
}
}
} 
\end{verbatim}
\vspace{-0.4cm}
&
\vspace{-0.4cm}
\begin{verbatim}
SELECT ?d WHERE { 
GRAPH <v1> { 
?d a qb:DimensionProperty. 
} 
FILTER NOT EXISTS { GRAPH <v2> { 
?d a qb:DimensionProperty. 
}
}
} 
\end{verbatim}
\vspace{-0.4cm}
\\ \hline

$\chng^+$ &
\isDimension{D}
 &
$\emptyset$ \\ \hline

$\chng^-$ &
$\emptyset$  &
\isDimension{D}
 \\ \hline

$\cond_{old}$ &
--  &
-- \\ \hline
$\cond_{new}$ &
--  &
-- \\ \hline

\end{tabular}
\end{footnotesize}
\newline
\end{table*}


\begin{table*}
\begin{footnotesize}
\begin{tabular}{|p{2cm}|p{6.5cm}|p{6.5cm}|}
\hline
Change &
\textbf{\op{Attach\_Datatype\_to\_Dimension($d,t$)}} &
\textbf{\op{Detach\_Datatype\_from\_Dimension($d,t$)}} \\ \hline

Intuition &
Associate a datatype with an existing dimension &
Disassociate a datatype from an existing dimension \\ \hline

Parameters &
$d$ = The dimension in which the datatype is attached \newline
$t$ = The datatype which is attached to dimension
&
$d$ = The dimension in which the datatype is detached \newline
$t$ = The datatype which is detached from dimension
\\ \hline

SPARQL used for detection &
\vspace{-0.4cm}
\begin{verbatim}SELECT ?d ?t WHERE { 
GRAPH <v2> { 
?d a qb:DimensionProperty. 
?d rdfs:range ?t. 
} 
FILTER NOT EXISTS { GRAPH <v1> { 
?d rdfs:range ?t. 
}
}
}
\end{verbatim} 
\vspace{-0.4cm}
&
\vspace{-0.4cm}
\begin{verbatim}SELECT ?d ?t WHERE { 
GRAPH <v1> { 
?d a qb:DimensionProperty. 
?d rdfs:range ?t. 
} 
FILTER NOT EXISTS { GRAPH <v2> { 
?d rdfs:range ?t. 
}
}
}
\end{verbatim}  
\vspace{-0.4cm}
\\ \hline

$\chng^+$ &
\hasRange{d}{t} &
$\emptyset$ \\ \hline

$\chng^-$ &
$\emptyset$  &
\hasRange{d}{t}
 \\ \hline

$\cond_{old}$ &
--  &
\uu{\isDimension{d}}  \\ \hline
$\cond_{new}$ &
\uu{\isDimension{d}}  &
-- \\ \hline

\end{tabular}
\end{footnotesize}
\newline
\end{table*}


\begin{table*}
\begin{footnotesize}
\begin{tabular}{|p{2cm}|p{6.5cm}|p{6.5cm}|}
\hline
Change &
\textbf{\op{Attach\_Attr\_to\_Dimension($d,attr$)}} &
\textbf{\op{Detach\_Attr\_from\_Dimension($d,attr$)}} \\ \hline

Intuition &
Associate an attribute property to a dimension &
Disassociate an attribute property from a dimension \\ \hline

Parameters &
$d$ = The dimension in which the atribute is attached \newline
$attr$ = The attribute which is attached to dimension
&
$d$ = The dimension in which the attribute is detached \newline
$attr$ = The attribute which is detached from dimension
\\ \hline

SPARQL used for detection &
\vspace{-0.4cm}
\begin{verbatim}
SELECT ?d ?attr WHERE { 
GRAPH <v2> {  
?d a qb:DimensionProperty. 
?d qb:attribute ?attr. 
} 
FILTER NOT EXISTS { GRAPH <v1> { 
?d qb:attribute ?attr. 
}
}
}
\end{verbatim} 
\vspace{-0.4cm} &
\vspace{-0.4cm}
\begin{verbatim}
SELECT ?d ?attr WHERE { 
GRAPH <v1> {  
?d a qb:DimensionProperty. 
?d qb:attribute ?attr. 
} 
FILTER NOT EXISTS { GRAPH <v2> { 
?d qb:attribute ?attr. 
}
}
}
\end{verbatim} 
\vspace{-0.4cm} \\ \hline

$\chng^+$ &
\hasAttribute{d}{attr} &
$\emptyset$ \\ \hline

$\chng^-$ &
$\emptyset$  &
\hasAttribute{d}{attr}
\\ \hline

$\cond_{old}$ &
--  &
\uu{\isDimension{d}}\\ \hline
$\cond_{new}$ &
\uu{\isDimension{d}}  &
-- \\ \hline

\end{tabular}
\end{footnotesize}
\newline
\end{table*}


\begin{table*}[h]
\begin{footnotesize}
\begin{tabular}{|p{2cm}|p{6.5cm}|p{6.5cm}|}
\hline
Change &
\textbf{\op{Add\_Observation($o$)}} &
\textbf{\op{Delete\_Observation($o$)}} \\ \hline

Intuition &
Add an observation entity &
Delete an observation entity \\ \hline

Parameters &
$o$ = The observation which is added
&
$o$ = The observation which is deleted \newline
\\ \hline

SPARQL used for detection &
\vspace{-0.4cm}
\begin{verbatim}
SELECT ?o WHERE { 
GRAPH <v2> { 
?o a qb:Observation. 
} 
FILTER NOT EXISTS { GRAPH <v1> { 
?o a qb:Observation. 
}
}
}
\end{verbatim}  
\vspace{-0.4cm} &

\vspace{-0.4cm}
\begin{verbatim}
SELECT ?o WHERE { 
GRAPH <v1> { 
?o a qb:Observation. 
} 
FILTER NOT EXISTS { GRAPH <v2> { 
?o a qb:Observation. 
}
}
}
\end{verbatim}  
\vspace{-0.4cm}  \\ \hline

$\chng^+$ &
\isObservation{o} &
$\emptyset$ \\ \hline

$\chng^-$ &
$\emptyset$  &
\isObservation{o}
\\ \hline

$\cond_{old}$ &
--  &
-- \\ \hline
$\cond_{new}$ &
--  &
-- \\ \hline

\end{tabular}
\end{footnotesize}
\newline
\end{table*}


\begin{table*}[h]
\begin{footnotesize}
\begin{tabular}{|p{2cm}|p{6.5cm}|p{6.5cm}|}
\hline
Change &
\textbf{\op{Attach\_Observation\_to\_FT($o,ft$)}} &
\textbf{\op{Detach\_Observation\_from\_FT($o,ft$)}} \\ \hline

Intuition &
Associate an observation to a fact table &
Disassociate an observation from a fact table \\ \hline

Parameters &
$o$ = The observation which is attached to fact table \newline
$ft$ = The fact table
&
$o$ = The observation which is detached from fact table \newline
$ft$ = The fact table
\\ \hline

SPARQL used for detection &
\vspace{-0.4cm}
\begin{verbatim}
SELECT ?o ?ft WHERE { 
GRAPH <v2> {  
?o qb:dataSet ?ds. 
?ds qb:structure ?ft. 
} 
FILTER NOT EXISTS { GRAPH <v1> { 
?o qb:dataSet ?ds. 
?ds qb:structure ?ft.  
}
}
}

\end{verbatim} 
\vspace{-0.4cm}
&
\vspace{-0.4cm}
\begin{verbatim}
SELECT ?o ?ft WHERE { 
GRAPH <v1> {  
?o qb:dataSet ?ds. 
?ds qb:structure ?ft. 
} 
FILTER NOT EXISTS { GRAPH <v2> { 
?o qb:dataSet ?ds. 
?ds qb:structure ?ft.   
}
}
}
\end{verbatim}  
\vspace{-0.4cm}
\\ \hline

$\chng^+$ &
\hasDataset{o}{ds},
\hasStructure{ds}{ft}
 &
$\emptyset$ \\ \hline

$\chng^-$ &
$\emptyset$  &
\hasDataset{o}{ds},
\hasStructure{ds}{ft}
\\ \hline

$\cond$ &
-- &
-- \\ \hline

\end{tabular}
\end{footnotesize}
\newline
\end{table*}


\begin{table*}[h]
\begin{footnotesize}
\begin{tabular}{|p{2cm}|p{6.5cm}|p{6.5cm}|}
\hline
Change &
\textbf{\op{Add\_Measure\_Value\_to\_Observation($o,m,v$)}} &
\textbf{\op{Delete\_Measure\_Value\_from\_Observation($o,m,v$)}} \\ \hline

Intuition &
Add value to a measure in a specific observation &
Delete a value from an observation \\ \hline

Parameters &
$o$ = The observation \newline
$m$ = The measure on which the observation refers \newline
$v$ = The value of the measure
&
$o$ = The observation \newline
$m$ = The measure on which the observation refers \newline
$v$ = The value of the measure
\\ \hline
SPARQL used for detection &
\vspace{-0.4cm}
\begin{verbatim}
SELECT ?o ?m ?v WHERE { 
GRAPH <v2> { 
?o qb:dataSet ?ds. 
?ds qb:structure ?ft. 
?ft qb:component ?cs. 
?cs qb:measure ?m. 
?m a qb:MeasureProperty. 
?m rdfs:range ?v. 
} 
FILTER NOT EXISTS { GRAPH <v1> { 
?o qb:dataSet ?ds. 
?ds qb:structure ?ft. 
?ft qb:component ?cs. 
?cs qb:measure ?m. 
?m a qb:MeasureProperty. 
?m rdfs:range ?v. 
}
}
}
\end{verbatim}  
\vspace{-0.4cm} &
\vspace{-0.4cm}
\begin{verbatim}
SELECT ?o ?m ?v WHERE { 
GRAPH <v1> { 
?o qb:dataSet ?ds. 
?ds qb:structure ?ft. 
?ft qb:component ?cs. 
?cs qb:measure ?m. 
?m a qb:MeasureProperty. 
?m rdfs:range ?v. 
} 
FILTER NOT EXISTS { GRAPH <v2> { 
?o qb:dataSet ?ds. 
?ds qb:structure ?ft. 
?ft qb:component ?cs. 
?cs qb:measure ?m. 
?m a qb:MeasureProperty. 
?m rdfs:range ?v. 
}
}
}
\end{verbatim}  
\vspace{-0.4cm}  \\ \hline

$\chng^+$ &
\hasDataset{o}{ds},
\hasStructure{ds}{ft},
\hasComponent{ft}{cs},
\hasMeasure{cs}{m},
\hasRange{m}{v}
&
$\emptyset$ \\ \hline

$\chng^-$ &
$\emptyset$  &
\hasDataset{o}{ds},
\hasStructure{ds}{ft},
\hasComponent{ft}{cs},
\hasMeasure{cs}{m},
\hasRange{m}{v}
\\ \hline

$\cond_{old}$ &
--  &
\uu{\isMeasure{m}}\\ \hline
$\cond_{new}$ &
\uu{\isMeasure{m}}  &
-- \\ \hline

\end{tabular}
\end{footnotesize}
\newline
\end{table*}


\begin{table*}[h]
\begin{footnotesize}
\begin{tabular}{|p{2cm}|p{6.5cm}|p{6.5cm}|}
\hline
Change &
\textbf{\op{Add\_Dimension\_Value\_to\_Observation($o,d,v$)}} &
\textbf{\op{Delete\_Dimension\_Value\_from\_Observation($o,d,v$)}} \\ \hline

Intuition &
Add value to a dimension in a specific observation &
Delete value from a dimension in an observation \\ \hline

Parameters &
$o$ = The observation \newline
$d$ = The dimension on which the observation refers \newline
$v$ = The value of the dimension
&
$o$ = The observation \newline
$d$ = The dimension on which the observation refers \newline
$v$ = The value of the dimension
\\ \hline

SPARQL used for detection &
\vspace{-0.4cm}
\begin{verbatim}
SELECT ?o ?d ?v WHERE { 
GRAPH <v2> { 
?o qb:dataSet ?ds. 
?ds qb:structure ?ft. 
?ft qb:component ?cs. 
?cs qb:dimension ?d. 
?d a qb:DimensionProperty. 
?d rdfs:range ?v. 
} 
FILTER NOT EXISTS { GRAPH <v1> { 
?o qb:dataSet ?ds. 
?ds qb:structure ?ft. 
?ft qb:component ?cs. 
?cs qb:dimension ?d. 
?d a qb:DimensionProperty. 
?d rdfs:range ?v. 
}
}
}
\end{verbatim}  
\vspace{-0.4cm} &
\vspace{-0.4cm}
\begin{verbatim}
SELECT ?o ?d ?v WHERE { 
GRAPH <v1> { 
?o qb:dataSet ?ds. 
?ds qb:structure ?ft. 
?ft qb:component ?cs. 
?cs qb:dimension ?d. 
?d a qb:DimensionProperty. 
?d rdfs:range ?v. 
} 
FILTER NOT EXISTS { GRAPH <v2> { 
?o qb:dataSet ?ds. 
?ds qb:structure ?ft. 
?ft qb:component ?cs. 
?cs qb:dimension ?d. 
?d a qb:DimensionProperty.  
?d rdfs:range ?v. 
}
}
}
\end{verbatim}  
\vspace{-0.4cm}  \\ \hline

$\chng^+$ &
\hasDataset{o}{ds},
\hasStructure{ds}{ft},
\hasComponent{ft}{cs},
\hasDimension{cs}{d},
\hasRange{d}{v}
&
$\emptyset$ \\ \hline

$\chng^-$ &
$\emptyset$  &
\hasDataset{o}{ds},
\hasStructure{ds}{ft},
\hasComponent{ft}{cs},
\hasDimension{cs}{d},
\hasRange{d}{v}
\\ \hline

$\cond_{old}$ &
--  &
\uu{\isDimension{d}}\\ \hline
$\cond_{new}$ &
\uu{\isDimension{d}}  &
-- \\ \hline

\end{tabular}
\end{footnotesize}
\newline
\end{table*}


\begin{table}[h]
\begin{footnotesize}
\begin{tabular}{|p{2cm}|p{6.5cm}|p{6.5cm}|}
\hline
Change &
\textbf{\op{Add\_Codelist($c$)}} &
\textbf{\op{Delete\_Codelist($c$)}} \\ \hline

Intuition &
Add a new codelist  &
Delete a codelist \\ \hline

Parameters &
$c$ = The added codelist &
$c$ = The deleted codelist\\ \hline

SPARQL used for detection &
\vspace{-0.4cm}
\begin{verbatim}
SELECT ?c WHERE { 
GRAPH <v2> {  
?c a skos:ConceptScheme. 
} 
FILTER NOT EXISTS { GRAPH <v1> { 
?c a skos:ConceptScheme. 
}
}
}
\end{verbatim}  
\vspace{-0.4cm} &
\vspace{-0.4cm}
\begin{verbatim}
SELECT ?c WHERE { 
GRAPH <v1> {  
?c a skos:ConceptScheme. 
} 
FILTER NOT EXISTS { GRAPH <v2> { 
?c a skos:ConceptScheme. 
}
}
}
\end{verbatim}  
\vspace{-0.4cm}  \\ \hline

$\chng^+$ &
\isCodelist{c} &
$\emptyset$ \\ \hline

$\chng^-$ &
$\emptyset$  &
\isCodelist{c}
\\ \hline

$\cond_{old}$ &
--  &
-- \\ \hline
$\cond_{new}$ &
--  &
-- \\ \hline

\end{tabular}
\end{footnotesize}
\newline
\end{table}


\begin{table*}[h]
\begin{footnotesize}
\begin{tabular}{|p{2cm}|p{6.5cm}|p{6.5cm}|}
\hline
Change &
\textbf{\op{Add\_Hierarchy($h$)}} &
\textbf{\op{Delete\_Hierarchy($h$)}} \\ \hline

Intuition &
Add a new hierarchy  &
Delete an hierarchy \\ \hline

Parameters &
$h$ = The added hierarchy &
$h$ = The deleted hierarchy\\ \hline

SPARQL used for detection &
\vspace{-0.4cm}
\begin{verbatim}
SELECT ?h WHERE { 
GRAPH <v2> { 
?h a qb:HierarchicalCodeList. 
FILTER NOT EXISTS 
{ ?h a skos:ConceptScheme. } 
} 
FILTER NOT EXISTS { GRAPH <v1> { 
?h a qb:HierarchicalCodeList. 
FILTER NOT EXISTS 
{ ?h a skos:ConceptScheme. } 
}
}
}
\end{verbatim}
\vspace{-0.4cm} &
\vspace{-0.4cm}
\begin{verbatim}
SELECT ?h WHERE { 
GRAPH <v1> { 
?h a qb:HierarchicalCodeList. 
FILTER NOT EXISTS 
{ ?h a skos:ConceptScheme. } 
} 
FILTER NOT EXISTS 
{ GRAPH <v2> { 
?h a qb:HierarchicalCodeList. 
FILTER NOT EXISTS 
{ ?h a skos:ConceptScheme. } 
}
}
}
\end{verbatim}
\vspace{-0.4cm}  \\ \hline

$\chng^+$ &
\isHierarchy{h}
 &
$\emptyset$ \\ \hline

$\chng^-$ &
$\emptyset$  &
\isHierarchy{h}
 \\ \hline

$\cond_{old}$ &
--  &
\begin{verbatim}FILTER NOT EXISTS 
{ ?h a skos:ConceptScheme. } \end{verbatim}  \\ \hline
$\cond_{new}$ &
\begin{verbatim}FILTER NOT EXISTS 
{ ?h a skos:ConceptScheme. } \end{verbatim}  &
-- \\ \hline

\end{tabular}
\end{footnotesize}
\newline
\end{table*}


\begin{table}[h]
\begin{footnotesize}
\begin{tabular}{|p{2cm}|p{6.5cm}|p{6.5cm}|}
\hline
Change &
\textbf{\op{Attach\_Codelist\_to\_Dimension($d,c$)}} &
\textbf{\op{Detach\_Codelist\_from\_Dimension($d,c$)}} \\ \hline

Intuition &
Assign a codelist to a dimension &
Disassociate a codelist from a dimension \\ \hline

Parameters &
$d$ = The dimension in which the codelist is attached \newline
$c$ = The codelist which is attached to dimension
&
$d$ = The dimension in which the codelist is detached \newline
$c$ = The codelist which is detached from dimension
\\ \hline

SPARQL used for detection &
\vspace{-0.4cm}
\begin{verbatim}
SELECT ?d ?c WHERE { 
GRAPH <v2> {  
?d qb:codeList ?c. 
} 
FILTER NOT EXISTS { GRAPH <v1> { 
?d qb:codeList ?c. 
}
}
}
\end{verbatim}  
\vspace{-0.4cm} &
\vspace{-0.4cm}
\begin{verbatim}
SELECT ?d ?c WHERE { 
GRAPH <v1> {  
?d qb:codeList ?c. 
} 
FILTER NOT EXISTS { GRAPH <v2> { 
?d qb:codeList ?c. 
}
}
}
\end{verbatim}  
\vspace{-0.4cm}  \\ \hline

$\chng^+$ &
\hasCodelist{d}{c} &
$\emptyset$ \\ \hline

$\chng^-$ &
$\emptyset$  &
\hasCodelist{d}{c}
\\ \hline

$\cond_{old}$ &
--  &
-- \\ \hline
$\cond_{new}$ &
--  &
-- \\ \hline

\end{tabular}
\end{footnotesize}
\newline
\end{table}


\begin{table*}[h]
\begin{footnotesize}
\begin{tabular}{|p{2cm}|p{6.5cm}|p{6.5cm}|}
\hline
Change &
\textbf{\op{Attach\_Hierarchy\_to\_Dimension($d,h$)}} &
\textbf{\op{Detach\_Hierarchy\_from\_Dimension($d,h$)}} \\ \hline

Intuition &
Associate an hierarchy to a dimension &
Disassociate an hierarchy from a dimension \\ \hline

Parameters &
$d$ = The dimension in which the hierarchy is attached \newline
$h$ = The hierarchy which is attached to dimension
&
$d$ = The dimension in which the hierarchy is detached \newline
$h$ = The hierarchy which is detached from dimension
\\ \hline

SPARQL used for detection &
\vspace{-0.4cm}
\begin{verbatim}
SELECT ?d ?h WHERE { 
GRAPH <v2> {  
?d qb:codeList ?h. 
?h a qb:HierarchicalCodeList. 
FILTER NOT EXISTS 
{ ?h a skos:ConceptScheme. } 
} 
FILTER NOT EXISTS { GRAPH <v1> { 
?d qb:codeList ?h. 
}
}
}
\end{verbatim}
\vspace{-0.4cm} &
\vspace{-0.4cm}
\begin{verbatim}
SELECT ?d ?h WHERE { 
GRAPH <v1> {  
?d qb:codeList ?h. 
?h a qb:HierarchicalCodeList. 
FILTER NOT EXISTS 
{ ?h a skos:ConceptScheme. } 
} 
FILTER NOT EXISTS { GRAPH <v2> { 
?d qb:codeList ?h. 
}
}
}
\end{verbatim}
\vspace{-0.4cm} \\ \hline

$\chng^+$ &
\hasCodelist{d}{h} &
$\emptyset$ \\ \hline

$\chng^-$ &
$\emptyset$  &
\hasCodelist{d}{h}
\\ \hline

$\cond_{old}$ &
--  &
\uu{\isHierarchy{h}}  \\ \hline
$\cond_{new}$ &
\uu{\isHierarchy{h}}  &
-- \\ \hline

\end{tabular}
\end{footnotesize}
\newline
\end{table*}


\begin{table*}[h]
\begin{footnotesize}
\begin{tabular}{|p{2cm}|p{6.5cm}|p{6.5cm}|}
\hline
Change &
\textbf{\op{Add\_Instance($i$)}} &
\textbf{\op{Delete\_Instance($i$)}} \\ \hline

Intuition &
Add a new instance &
Delete an instance \\ \hline

Parameters &
$i$ = The added instance &
$i$ = The deleted instance\\ \hline

SPARQL used for detection &
\vspace{-0.4cm}
\begin{verbatim}
SELECT ?i WHERE { 
GRAPH <v2> {  
?i a skos:Concept. 
} 
FILTER NOT EXISTS { GRAPH <v1> { 
?i a skos:Concept. 
}
}
}
\end{verbatim}
\vspace{-0.4cm} &
\vspace{-0.4cm}
\begin{verbatim}
SELECT ?i WHERE { 
GRAPH <v1> {  
?i a skos:Concept. 
} 
FILTER NOT EXISTS { GRAPH <v2> { 
?i a skos:Concept. 
}
}
}
\end{verbatim}
\vspace{-0.4cm}  \\ \hline

$\chng^+$ &
\isInstance{i}
 &
$\emptyset$ \\ \hline

$\chng^-$ &
$\emptyset$  &
\isInstance{i}
 \\ \hline

$\cond_{old}$ &
--  &
-- \\ \hline
$\cond_{new}$ &
--  &
-- \\ \hline

\end{tabular}
\end{footnotesize}
\newline
\end{table*}


\begin{table*}[h]
\begin{footnotesize}
\begin{tabular}{|p{2cm}|p{6.5cm}|p{6.5cm}|}
\hline
Change &
\textbf{\op{Attach\_Instance\_to\_Codelist($c,i$)}} &
\textbf{\op{Detach\_Instance\_from\_Codelist($c,i$)}} \\ \hline

Intuition &
Associate a new instance with a codelist  &
Disassociate an instance from a codelist \\ \hline

Parameters &
$c$ = The codelist in which the instance is attached \newline
$i$ = The instance which is attached to codelist
&
$c$ = The codelist in which the instance is attached \newline
$i$ = The instance which is attached to codelist
\\ \hline

SPARQL used for detection &
\vspace{-0.4cm}
\begin{verbatim}

SELECT ?c ?i WHERE { 
GRAPH <v2> {  
?i a skos:ConceptScheme. 
?c skos:inScheme ?i. 
} 
FILTER NOT EXISTS { GRAPH <v1> { 
?c skos:inScheme ?i. 
}
}
}
\end{verbatim}
\vspace{-0.4cm} &
\vspace{-0.4cm}
\begin{verbatim}

SELECT ?c ?i WHERE { 
GRAPH <v1> {  
?i a skos:ConceptScheme. 
?c skos:inScheme ?i. 
} 
FILTER NOT EXISTS { GRAPH <v2> { 
?c skos:inScheme ?i. 
}
}
}
\end{verbatim}
\vspace{-0.4cm}  \\ \hline

$\chng^+$ &
\inScheme{i}{c} &
$\emptyset$ \\ \hline

$\chng^-$ &
$\emptyset$  &
\inScheme{i}{c}
\\ \hline

$\cond_{old}$ &
--  &
\uu{\isCodelist{c}}  \\ \hline
$\cond_{new}$ &
\uu{\isCodelist{c}}  &
-- \\ \hline

\end{tabular}
\end{footnotesize}
\newline
\end{table*}


\begin{table*}[h]
\begin{footnotesize}
\begin{tabular}{|p{2cm}|p{6.5cm}|p{6.5cm}|}
\hline
Change &
\textbf{\op{Attach\_Instance\_to\_Hierarchy($h,i$)}} &
\textbf{\op{Detach\_Instance\_from\_Hierarchy($h,i$)}} \\ \hline

Intuition &
Associate a new instance with a hierarchy  &
Disassociate a new instance from a hierarchy  \\ \hline

Parameters &
$h$ = The hierarchy in which the instance is attached \newline
$i$ = The instance which is attached to hierarchy
&
$h$ = The hierarchy in which the instance is attached \newline
$i$ = The instance which is attached to hierarchy
\\ \hline

SPARQL used for detection &
\vspace{-0.4cm}
\begin{verbatim}
SELECT ?h ?i WHERE { 
GRAPH <v2> {  
?h a qb:HierarchicalCodeList. 
?i skos:inScheme ?h. 
} 
FILTER NOT EXISTS { GRAPH <v1> { 
?i skos:inScheme ?h. 
}
}
}
\end{verbatim}
\vspace{-0.4cm} &
\vspace{-0.4cm}
\begin{verbatim}
SELECT ?h ?i WHERE { 
GRAPH <v1> {  
?h a qb:HierarchicalCodeList. 
?i skos:inScheme ?h. 
} 
FILTER NOT EXISTS { GRAPH <v2> { 
?i skos:inScheme ?h. 
}
}
}
\end{verbatim}
\vspace{-0.4cm}  \\ \hline

$\chng^+$ &
\inScheme{i}{h} &
$\emptyset$ \\ \hline

$\chng^-$ &
$\emptyset$  &
\inScheme{i}{h}
\\ \hline

$\cond_{old}$ &
--  &
\uu{\isHierarchy{h}}  \\ \hline
$\cond_{new}$ &
\uu{\isHierarchy{h}}  &
-- \\ \hline

\end{tabular}
\end{footnotesize}
\newline
\end{table*}


\begin{table*}[h]
\begin{footnotesize}
\begin{tabular}{|p{2cm}|p{6.5cm}|p{6.5cm}|}
\hline
Change &
\textbf{\op{Attach\_Instance\_to\_Parent($i,p$)}} &
\textbf{\op{Detach\_Instance\_to\_Parent($i,p$)}} \\ \hline

Intuition &
Associate an instance with its parent  &
Disassociate an instance from its parent \\ \hline

Parameters &
$i$ = The instance \newline
$p$ = The parent which is added to instance
&
$i$ = The instance \newline
$p$ = The parent which is deleted from instance
\\ \hline

SPARQL used for detection &
\vspace{-0.4cm}
\begin{verbatim}
SELECT ?i ?p WHERE { 
GRAPH <v2> {  
?i skos:broaderTransitive ?p.  
} 
FILTER NOT EXISTS { GRAPH <v1> {  
?i skos:broaderTransitive ?p.  
}
}
}
\end{verbatim}
\vspace{-0.4cm}
 &
\vspace{-0.4cm}
\begin{verbatim}
SELECT ?i ?p WHERE { 
GRAPH <v1> {  
?i skos:broaderTransitive ?p.  
} 
FILTER NOT EXISTS { GRAPH <v2> {  
?i skos:broaderTransitive ?p.  
}
}
}
\end{verbatim}
\vspace{-0.4cm}  \\ \hline

$\chng^+$ &
\hasParent{i}{p} &
$\emptyset$ \\ \hline

$\chng^-$ &
$\emptyset$  &
\hasParent{i}{p} 
\\ \hline

$\cond_{old}$ &
--  &
-- \\ \hline
$\cond_{new}$ &
--  &
-- \\ \hline

\end{tabular}
\end{footnotesize}
\newline
\end{table*}



\begin{table*}[h]
\begin{footnotesize}
\begin{tabular}{|p{2cm}|p{6.5cm}|p{6.5cm}|}
\hline
Change &
\textbf{\op{Add\_Measure($m$)}} &
\textbf{\op{Delete\_Measure($m$)}} \\ \hline

Intuition &
Add a new measure  &
Delete a measure \\ \hline

Parameters &
$m$ = The measure which is added &
$m$ = The measure which is deleted\\ \hline

SPARQL used for detection &
\vspace{-0.4cm}
\begin{verbatim}
SELECT ?m WHERE { 
GRAPH <v2> {  
?m a qb:MeasureProperty. 
} 
FILTER NOT EXISTS { GRAPH <v1> { 
?m a qb:MeasureProperty. 
}
}
}
\end{verbatim}
\vspace{-0.4cm} &
\vspace{-0.4cm}
\begin{verbatim}
SELECT ?m WHERE { 
GRAPH <v1> {  
?m a qb:MeasureProperty. 
} 
FILTER NOT EXISTS { GRAPH <v2> { 
?m a qb:MeasureProperty. 
}
}
}
\end{verbatim}
\vspace{-0.4cm}  \\ \hline

$\chng^+$ &
\isMeasure{m}
 &
$\emptyset$ \\ \hline

$\chng^-$ &
$\emptyset$  &
\isMeasure{m}
 \\ \hline

$\cond_{old}$ &
--  &
-- \\ \hline
$\cond_{new}$ &
--  &
-- \\ \hline

\end{tabular}
\end{footnotesize}
\newline
\end{table*}


\begin{table*}[h]
\begin{footnotesize}
\begin{tabular}{|p{2cm}|p{6.5cm}|p{6.5cm}|}
\hline
Change &
\textbf{\op{Attach\_Type\_to\_Measure($t,m$)}} &
\textbf{\op{Detach\_Type\_from\_Measure($t,m$)}} \\ \hline

Intuition &
Associate a new datatype to a measure  &
Disassociate a datatype from a measure \\ \hline

Parameters &
$t$ = The added type to measure \newline
$m$ = The measure in which the type is added
&
$t$ = The deleted type from measure \newline
$m$ = The measure in which the type is deleted
\\ \hline

SPARQL used for detection &
\vspace{-0.4cm}
\begin{verbatim}
SELECT ?m ?t WHERE { 
GRAPH <v2> { 
?m a qb:MeasureProperty. 
?m rdfs:range ?t. 
}  
FILTER NOT EXISTS { GRAPH <v1> {  
?m rdfs:range ?t. 
}
}
}
\end{verbatim}
\vspace{-0.4cm} &
\vspace{-0.4cm}
\begin{verbatim}
SELECT ?m ?t WHERE { 
GRAPH <v1> { 
?m a qb:MeasureProperty. 
?m rdfs:range ?t. 
}  
FILTER NOT EXISTS { GRAPH <v2> {  
?m rdfs:range ?t. 
}
}
}
\end{verbatim}
\vspace{-0.4cm}   \\ \hline

$\chng^+$ &
\hasRange{m}{t} &
$\emptyset$ \\ \hline

$\chng^-$ &
$\emptyset$  &
\hasRange{m}{t}
\\ \hline

$\cond_{old}$ &
--  &
\uu{\isMeasure{m}}  \\ \hline
$\cond_{new}$ &
\uu{\isMeasure{m}}  &
-- \\ \hline

\end{tabular}
\end{footnotesize}
\newline
\end{table*}



\begin{table*}[h]
\begin{footnotesize}
\begin{tabular}{|p{2cm}|p{6.5cm}|p{6.5cm}|}
\hline
Change &
\textbf{\op{Add\_Fact\_Table($ft$)}} &
\textbf{\op{Delete\_Fact\_Table($ft$)}} \\ \hline

Intuition &
Add a new fact table  &
Delete a fact table \\ \hline

Parameters &
$ft$ = The fact table is added &
$ft$ = The fact table which is deleted\\ \hline

SPARQL used for detection &
\vspace{-0.4cm}
\begin{verbatim}
SELECT ?ft WHERE { 
GRAPH <v2> { 
?ft a qb:DataStructureDefinition. 
} 
FILTER NOT EXISTS { GRAPH <v1> { 
?ft a qb:DataStructureDefinition. 
}
}
}
\end{verbatim}
\vspace{-0.4cm} &
\vspace{-0.4cm}
\begin{verbatim}
SELECT ?ft WHERE { 
GRAPH <v1> { 
?ft a qb:DataStructureDefinition. 
} 
FILTER NOT EXISTS { GRAPH <v2> { 
?ft a qb:DataStructureDefinition. 
}
}
}
\end{verbatim}
\vspace{-0.4cm}  \\ \hline

$\chng^+$ &
\isFT{ft} &
$\emptyset$ \\ \hline

$\chng^-$ &
$\emptyset$  &
\isFT{ft}
\\ \hline

$\cond_{old}$ &
--  &
-- \\ \hline
$\cond_{new}$ &
--  &
-- \\ \hline

\end{tabular}
\end{footnotesize}
\newline
\end{table*}

\clearpage
\begin{table*}[h]

\begin{footnotesize}
\begin{tabular}{|p{2cm}|p{6.5cm}|p{6.5cm}|}
\hline
Change &
\textbf{\op{Attach\_Measure\_to\_Fact\_Table($m,ft$)}} &
\textbf{\op{Detach\_Measure\_from\_Fact\_Table($m,ft$)}} \\ \hline

Intuition &
Associate a measure property to fact table  &
Disassociate a measure property from a fact table \\ \hline

Parameters &
$m$ = The added measure to fact table \newline
$ft$ = The fact table in which the measure is added
&
$m$ = The deleted measure to fact table \newline
$ft$ = The fact table in which the measure is deleted
\\ \hline

SPARQL used for detection &
\vspace{-0.4cm}
\begin{verbatim}
SELECT ?ft ?m WHERE { 
GRAPH <v2> {  
?ft qb:component ?cs. 
?cs qb:measure ?m. 
} 
FILTER NOT EXISTS { GRAPH <v1> { 
?ft qb:component ?cs. 
?cs qb:measure ?m. 
}
}
}
\end{verbatim}
\vspace{-0.4cm} &
\vspace{-0.4cm}
\begin{verbatim}
SELECT ?ft ?m WHERE { 
GRAPH <v1> {  
?ft qb:component ?cs. 
?cs qb:measure ?m. 
} 
FILTER NOT EXISTS { GRAPH <v2> { 
?ft qb:component ?cs. 
?cs qb:measure ?m. 
}
}
}
\end{verbatim}
\vspace{-0.4cm}  \\ \hline

$\chng^+$ &
\hasComponent{ft}{cs},
\hasMeasure{cs}{m} &
$\emptyset$ \\ \hline

$\chng^-$ &
$\emptyset$  &
\hasComponent{ft}{cs},
\hasMeasure{cs}{m}
\\ \hline

$\cond_{old}$ &
--  &
-- \\ \hline
$\cond_{new}$ &
--  &
-- \\ \hline

\end{tabular}
\end{footnotesize}
\newline

\end{table*}


\begin{table*}[h]
\begin{footnotesize}
\begin{tabular}{|p{2cm}|p{6.5cm}|p{6.5cm}|}
\hline
Change &
\textbf{\op{Attach\_Dimension\_to\_Fact\_Table($d,fT$)}} &
\textbf{\op{Detach\_Dimension\_from\_Fact\_Table($d,ft$)}} \\ \hline

Intuition &
Associate a dimension property to fact table  &
Disassociate a dimension property from a fact table  \\ \hline

Parameters &
$d$ = The added dimension to fact table \newline
$ft$ = The fact table in which the dimension is added
&
$d$ = The deleted dimension to fact table \newline
$ft$ = The fact table in which the dimension is deleted
\\ \hline

SPARQL used for detection &
\vspace{-0.4cm}
\begin{verbatim}
SELECT ?d ?ft WHERE { 
GRAPH <v2> {  
?ft qb:component ?cs. 
?cs qb:dimension ?d. 
} 
FILTER NOT EXISTS { GRAPH <v1> { 
?ft qb:component ?cs. 
?cs qb:dimension ?d.
}
}
}
\end{verbatim}
\vspace{-0.4cm} &
\vspace{-0.4cm}
\begin{verbatim}
SELECT ?d ?ft WHERE { 
GRAPH <v1> {  
?ft qb:component ?cs. 
?cs qb:dimension ?d. 
} 
FILTER NOT EXISTS { GRAPH <v2> { 
?ft qb:component ?cs. 
?cs qb:dimension ?d.
}
}
}
\end{verbatim}
\vspace{-0.4cm}  \\ \hline

$\chng^+$ &
\hasComponent{ft}{cs},
\hasDimension{cs}{d} &
$\emptyset$ \\ \hline

$\chng^-$ &
$\emptyset$  &
\hasComponent{ft}{cs},
\hasDimension{cs}{d}
\\ \hline

$\cond_{old}$ &
--  &
-- \\ \hline
$\cond_{new}$ &
--  &
-- \\ \hline

\end{tabular}
\end{footnotesize}
\newline
\end{table*}



\begin{table*}[h]
\begin{footnotesize}
\begin{tabular}{|p{2cm}|p{6.5cm}|p{6.5cm}|}
\hline
Change &
\textbf{\op{Add\_Attribute($attr$)}} &
\textbf{\op{Delete\_Attribute($attr$)}} \\ \hline

Intuition &
Add a new attribute  &
Delete an attribute \\ \hline

Parameters &
$attr$ = The attribute which is added &
$attr$ = The attribute which is deleted\\ \hline

SPARQL used for detection &
\vspace{-0.4cm}
\begin{verbatim}
SELECT ?attr WHERE { 
GRAPH <v2> { 
?attr a qb:AttributeProperty. 
} 
FILTER NOT EXISTS { GRAPH <v1> { 
?attr a qb:AttributeProperty. 
}
}
}
\end{verbatim}
\vspace{-0.4cm} &
\vspace{-0.4cm}
\begin{verbatim}
SELECT ?attr WHERE { 
GRAPH <v1> { 
?attr a qb:AttributeProperty. 
} 
FILTER NOT EXISTS { GRAPH <v2> { 
?attr a qb:AttributeProperty. 
}
}
}
\end{verbatim}
\vspace{-0.4cm}  \\ \hline

$\chng^+$ &
\isAttribute{attr}
 &
$\emptyset$ \\ \hline

$\chng^-$ &
$\emptyset$  &
\isAttribute{attr}
 \\ \hline

$\cond_{old}$ &
--  &
-- \\ \hline
$\cond_{new}$ &
--  &
-- \\ \hline

\end{tabular}
\end{footnotesize}
\newline
\end{table*}


\begin{table*}[h]
\begin{footnotesize}
\begin{tabular}{|p{2cm}|p{6.5cm}|p{6.5cm}|}
\hline
Change &
\textbf{\op{Attach\_Attr\_to\_Measure($attr,m$)}} &
\textbf{\op{Detach\_Attr\_from\_Measure($attr,m$)}} \\ \hline

Intuition &
Associate an attribute with an existing measure  &
Disassociate an attribute from a measure \\ \hline

Parameters &
$attr$ = The attribute which is attached to measure 
$m$ = The measure which is attached
&
$attr$ = The attribute which is detached from measure
\newline
$m$ = The measure in which is the measure is detatched\\ \hline

SPARQL used for detection &
\vspace{-0.4cm}
\begin{verbatim}
SELECT ?attr ?m WHERE { 
GRAPH <v2> {  
?m a qb:MeasureProperty. 
?m qb:attribute ?attr. 
} 
FILTER NOT EXISTS { GRAPH <v1> { 
?m qb:attribute ?attr. 
}
}
}
\end{verbatim}
\vspace{-0.4cm} &
\vspace{-0.4cm}
\begin{verbatim}
SELECT ?attr ?m WHERE { 
GRAPH <v1> {  
?m a qb:MeasureProperty. 
?m qb:attribute ?attr. 
} 
FILTER NOT EXISTS { GRAPH <v2> { 
?m qb:attribute ?attr. 
}
}
}
\end{verbatim}
\vspace{-0.4cm}  \\ \hline

$\chng^+$ &
\hasAttribute{m}{attr}
 &
$\emptyset$ \\ \hline

$\chng^-$ &
$\emptyset$  &
\hasAttribute{m}{attr}
 \\ \hline

$\cond_{old}$ &
--  &
\uu{\isMeasure{m}}  \\ \hline
$\cond_{new}$ &
\uu{\isMeasure{m}}  &
-- \\ \hline

\end{tabular}
\end{footnotesize}
\newline
\end{table*}


\begin{table*}[h]
\begin{footnotesize}
\begin{tabular}{|p{2cm}|p{6.5cm}|p{6.5cm}|}
\hline
Change &
\textbf{\op{Attach\_Observation\_to\_Dataset($o,ds$)}} &
\textbf{\op{Detach\_Observation\_from\_Dataset($o,ds$)}} \\ \hline

Intuition &
Associate an observation with an existing dataset  &
Disassociate an observation from a dataset \\ \hline

Parameters &
$o$ = The observation which is attached to dataset \newline
$ds$ = The dataset which in which the observation is attached
&
$o$ = The observation which is detached from dataset
$ds$ = The dataset in which is the observation is detatched\\ \hline

SPARQL used for detection &
\vspace{-0.4cm}
\begin{verbatim}
SELECT ?o ?ds WHERE { 
GRAPH <v2> {
?o qb:dataSet ?ds. 
} 
FILTER NOT EXISTS { GRAPH <v1> { 
?o qb:dataSet ?ds.
}
}
}
\end{verbatim}
\vspace{-0.4cm} &
\vspace{-0.4cm}
\begin{verbatim}
SELECT ?o ?ds WHERE { 
GRAPH <v1> {
?o qb:dataSet ?ds. 
} 
FILTER NOT EXISTS { GRAPH <v2> { 
?o qb:dataSet ?ds.
}
}
}
\end{verbatim}
\vspace{-0.4cm}  \\ \hline

$\chng^+$ &
\hasDataset{o}{ds}
 &
$\emptyset$ \\ \hline

$\chng^-$ &
$\emptyset$  &
\hasDataset{o}{ds}
 \\ \hline

$\cond_{old}$ &
--  &
-- \\ \hline
$\cond_{new}$ &
--  &
-- \\ \hline

\end{tabular}
\end{footnotesize}
\newline
\end{table*}


\begin{table*}[h]
\begin{footnotesize}
\begin{tabular}{|p{2cm}|p{6.5cm}|p{6.5cm}|}
\hline
Change &
\textbf{\op{Add\_Inscheme($x,s$)}} &
\textbf{\op{Delete\_Inscheme($x,s$)}} \\ \hline

Intuition &
Add a scheme in a term  &
Delete a scheme from a term \\ \hline

Parameters &
$x$ = The term which is associated to the scheme \newline
$s$ = The associated scheme
&
$x$ = The term which is deleted from associated scheme \newline
$s$ = The associated scheme\\ \hline

SPARQL used for detection &
\vspace{-0.4cm}
\begin{verbatim}
SELECT ?x ?s WHERE { 
GRAPH <v2> {  
FILTER NOT EXISTS { {
?s rdf:type skos:ConceptScheme.} 
UNION  
{?s rdf:type 
qb:HierarchicalCodeList.} 
} 
?x skos:inScheme ?s. 
} 
FILTER NOT EXISTS { GRAPH <v1> { 
?x skos:inScheme  ?s. 
}
}
}
\end{verbatim}
\vspace{-0.4cm} &
\vspace{-0.4cm}
\begin{verbatim}
SELECT ?x ?s WHERE { 
GRAPH <v1> {  
FILTER NOT EXISTS { 
{?s rdf:type 
skos:ConceptScheme.} 
UNION  
{?s rdf:type 
qb:HierarchicalCodeList.} 
} 
?x skos:inScheme ?s. 
} 
FILTER NOT EXISTS { GRAPH <v2> { 
?x skos:inScheme ?s. 
}
}
}
\end{verbatim}
\vspace{-0.4cm}  \\ \hline

$\chng^+$ &
\inScheme{x}{s}
 &
$\emptyset$ \\ \hline

$\chng^-$ &
$\emptyset$  &
\inScheme{x}{s}
 \\ \hline

$\cond_{old}$ &
--  &
\begin{verbatim}
FILTER NOT EXISTS { 
{?s rdf:type skos:ConceptScheme.} 
UNION  
{?s rdf:type 
qb:HierarchicalCodeList.} 
} 
\end{verbatim}
\\ \hline
$\cond_{new}$ &
\begin{verbatim}
FILTER NOT EXISTS { 
{?s rdf:type 
skos:ConceptScheme.} 
UNION  
{?s rdf:type
 qb:HierarchicalCodeList.} 
} 
\end{verbatim} &
-- \\ \hline

\end{tabular}
\end{footnotesize}
\newline
\end{table*}



\begin{table}[h]
\begin{footnotesize}
\begin{tabular}{|p{2cm}|p{6.5cm}|p{6.5cm}|}
\hline
Change &
\textbf{\op{Add\_Label($s,o$)}} &
\textbf{\op{Delete\_Label($s,o$)}} \\ \hline

Intuition &
Add a new label &
Delete a label \\ \hline

Parameters &
$s$ = The subject in which the label is added \newline
$o$ = The added label
&
$s$ = The subject in which the label is deleted \newline
$o$ = The deleted label\\ \hline

SPARQL used for detection &
\vspace{-0.4cm}
\begin{verbatim}
SELECT ?s ?o WHERE { 
GRAPH <v2> { 
?s ?p ?o. 
filter (?p = rdfs:label).
} 
FILTER NOT EXISTS { GRAPH <v1> { 
?s ?p ?o 
}
}
}
\end{verbatim}
\vspace{-0.4cm} &
\vspace{-0.4cm}
\begin{verbatim}
SELECT ?s ?o WHERE { 
GRAPH <v1> { 
?s ?p ?o. 
filter (?p = rdfs:label).
} 
FILTER NOT EXISTS { GRAPH <v2> { 
?s ?p ?o 
}
}
}
\end{verbatim}
\vspace{-0.4cm}  \\ \hline

$\chng^+$ 
&
\triple{a}{\labell}{b}
&
$\emptyset$ \\ \hline

$\chng^-$ &
$\emptyset$  &
\triple{a}{\labell}{b} \\ \hline

$\cond_{old}$ &
-- &
--\\ \hline

$\cond_{new}$ &
--
&
--
\\ \hline

\end{tabular}
\end{footnotesize}
\newline
\end{table}


\begin{table*}[h]
\begin{footnotesize}
\begin{tabular}{|p{2cm}|p{7.5cm}|p{7.5cm}|}
\hline
Change &
\textbf{\op{Add\_Unknown\_Property($s,p,o$)}} &
\textbf{\op{Delete\_Unknown\_Property($s,p,o$)}} \\ \hline

Intuition &
Add a new (unknown) property with specified subject and object related &
Delete a property \\ \hline

Parameters &
$s$ = The subject in which the property is added \newline
$p$ = The property  \newline
$o$ = The object which is related with the subject via the property
&
$s$ = The subject in which the property is deleted \newline
$p$ = The property  \newline
$o$ = The object which is related with the subject via the property\\ \hline

SPARQL used for detection &
\vspace{-0.4cm}
\begin{verbatim}
SELECT ?s ?p ?o WHERE { 
GRAPH <v2> { 
{FILTER NOT EXISTS 
{?p rdfs:subPropertyOf rdfs:label}} 
UNION 
{FILTER (?p != rdfs:label).} 
?s ?p ?o. 
FILTER(?p != rdfs:label). 
FILTER(?p != rdfs:range). 
FILTER(?p != skos:inScheme). 
FILTER(?p != skos:broaderTransitive). 
FILTER(?p != qb:codeList). 
FILTER(?p != qb:component). 
FILTER(?p != qb:dimension). 
FILTER(?p != qb:measure). 
FILTER(?p != qb:attribute). 
FILTER(?p != qb:dataSet). 
FILTER(?p != qb:structure). 
FILTER(?p != rdf:type 
|| skos:Concept). 
FILTER (?p != rdf:type 
|| skos:ConceptScheme). 
FILTER (?p != rdf:type
|| qb:AttributeProperty). 
FILTER (?p != rdf:type 
|| qb:CodedProperty). 
FILTER (?p != rdf:type 
|| qb:DimensionProperty). 
FILTER (?p != rdf:type 
|| qb:DataStructureDefinition). 
FILTER (?p != rdf:type 
|| qb:HierarchicalCodeList). 
FILTER (?p != rdf:type 
|| qb:Observation). } 
FILTER NOT EXISTS 
{ GRAPH <v1> { 
?s ?p ?o. 
}
}
}
\end{verbatim}
\vspace{-0.4cm} &
\vspace{-0.4cm}
\begin{verbatim}
SELECT ?s ?p ?o WHERE { 
GRAPH <v1> { 
{FILTER NOT EXISTS 
{?p rdfs:subPropertyOf rdfs:label}} 
UNION 
{FILTER (?p != rdfs:label).} 
?s ?p ?o. 
FILTER(?p != rdfs:label). 
FILTER(?p != rdfs:range). 
FILTER(?p != skos:inScheme). 
FILTER(?p != skos:broaderTransitive). 
FILTER(?p != qb:codeList). 
FILTER(?p != qb:component). 
FILTER(?p != qb:dimension). 
FILTER(?p != qb:measure). 
FILTER(?p != qb:attribute). 
FILTER(?p != qb:dataSet). 
FILTER(?p != qb:structure). 
FILTER(?p != rdf:type 
|| skos:Concept). 
FILTER (?p != rdf:type 
|| skos:ConceptScheme). 
FILTER (?p != rdf:type
|| qb:AttributeProperty). 
FILTER (?p != rdf:type 
|| qb:CodedProperty). 
FILTER (?p != rdf:type 
|| qb:DimensionProperty). 
FILTER (?p != rdf:type 
|| qb:DataStructureDefinition). 
FILTER (?p != rdf:type 
|| qb:HierarchicalCodeList). 
FILTER (?p != rdf:type 
|| qb:Observation). } 
FILTER NOT EXISTS 
{ GRAPH <v1> { 
?s ?p ?o. 
}
}
}
\end{verbatim}
\vspace{-0.4cm} \\ \hline

$\chng^+$ &
\triple{s}{p}{o} 
 &
$\emptyset$ \\ \hline

$\chng^-$ &
$\emptyset$  &
\triple{s}{p}{o} 
 \\ \hline
\end{tabular}
\end{footnotesize}
\newline
\end{table*}

\begin{table*}[h]
\begin{footnotesize}
\begin{tabular}{|p{2cm}|p{6.5cm}|p{6.5cm}|}
\hline
$\cond_{old}$ &
--  &
\begin{verbatim}
{FILTER NOT EXISTS 
{?p rdfs:subPropertyOf rdfs:label}} 
UNION 
{FILTER (?p != rdfs:label).} 
?s ?p ?o. 
FILTER(?p != rdfs:label). 
FILTER(?p != rdfs:range). 
FILTER(?p != skos:inScheme). 
FILTER(?p != skos:broaderTransitive). 
FILTER(?p != qb:codeList). 
FILTER(?p != qb:component). 
FILTER(?p != qb:dimension). 
FILTER(?p != qb:measure). 
FILTER(?p != qb:attribute). 
FILTER(?p != qb:dataSet). 
FILTER(?p != qb:structure). 
FILTER(?p != rdf:type 
|| skos:Concept). 
FILTER (?p != rdf:type 
|| skos:ConceptScheme). 
FILTER (?p != rdf:type
|| qb:AttributeProperty). 
FILTER (?p != rdf:type 
|| qb:CodedProperty). 
FILTER (?p != rdf:type 
|| qb:DimensionProperty). 
FILTER (?p != rdf:type 
|| qb:DataStructureDefinition). 
FILTER (?p != rdf:type 
|| qb:HierarchicalCodeList). 
FILTER (?p != rdf:type 
|| qb:Observation). 
\end{verbatim} \\ \hline
$\cond_{new}$ &
\begin{verbatim}
{FILTER NOT EXISTS 
{?p rdfs:subPropertyOf rdfs:label}} 
UNION 
{FILTER (?p != rdfs:label).} 
?s ?p ?o. 
FILTER(?p != rdfs:label). 
FILTER(?p != rdfs:range). 
FILTER(?p != skos:inScheme). 
FILTER(?p != skos:broaderTransitive). 
FILTER(?p != qb:codeList). 
FILTER(?p != qb:component). 
FILTER(?p != qb:dimension). 
FILTER(?p != qb:measure). 
FILTER(?p != qb:attribute). 
FILTER(?p != qb:dataSet). 
FILTER(?p != qb:structure). 
FILTER(?p != rdf:type 
|| skos:Concept). 
FILTER (?p != rdf:type 
|| skos:ConceptScheme). 
FILTER (?p != rdf:type
|| qb:AttributeProperty). 
FILTER (?p != rdf:type 
|| qb:CodedProperty). 
FILTER (?p != rdf:type 
|| qb:DimensionProperty). 
FILTER (?p != rdf:type 
|| qb:DataStructureDefinition). 
FILTER (?p != rdf:type 
|| qb:HierarchicalCodeList). 
FILTER (?p != rdf:type 
|| qb:Observation). 
\end{verbatim}  &
-- \\ \hline

\end{tabular}
\end{footnotesize}
\newline
\end{table*}


\begin{table*}[h]
\begin{footnotesize}
\begin{tabular}{|p{2cm}|p{6.5cm}|p{6.5cm}|}
\hline
Change &
\textbf{\op{Add\_Generic\_Datatype($x,t$)}} &
\textbf{\op{Delete\_Generic\_Datatype($x,t$)}} \\ \hline

Intuition &
Add a data type to a given subject &
Delete a data type from a subject \\ \hline

Parameters &
$x$ = The subject in which the datatype is added \newline
$t$ = The added datatype &
$x$ = The subject in which the datatype is deleted  \newline
$t$ = The deleted datatype\\ \hline

SPARQL used for detection &
\vspace{-0.4cm}
\begin{verbatim}
SELECT ?x ?t WHERE { 
GRAPH <v2> {  
?x rdfs:range ?t.  
}  
FILTER NOT EXISTS { GRAPH <v1> {  
?x rdfs:range ?t.  
}
}
}
\end{verbatim}
\vspace{-0.4cm} &
\vspace{-0.4cm}
\begin{verbatim}
SELECT ?x ?t WHERE { 
GRAPH <v1> {  
?x rdfs:range ?t.  
}  
FILTER NOT EXISTS { GRAPH <v2> {  
?x rdfs:range ?t.  
}
}
}
\end{verbatim}
\vspace{-0.4cm}  \\ \hline

$\chng^+$ &
\hasRange{x}{t}
 &
$\emptyset$ \\ \hline

$\chng^-$ &
$\emptyset$  &
\hasRange{x}{t}
 \\ \hline

$\cond_{old}$ &
--  &
-- \\ \hline
$\cond_{new}$ &
--  &
-- \\ \hline

\end{tabular}
\end{footnotesize}
\newline
\end{table*}

\begin{table*}[h]
\begin{footnotesize}
\begin{tabular}{|p{2cm}|p{6.5cm}|p{6.5cm}|}
\hline
Change &
\textbf{\op{Add\_Generic\_Attribute($x,attr$)}} &
\textbf{\op{Delete\_Generic\_Attribute($x,attr$)}} \\ \hline

Intuition &
Add a generic attribute to a given subject&
Delete a generic attribute from a subject \\ \hline

Parameters &
$x$ = The subject in which the attribute is added \newline
$attr$ = The added attribute &
$x$ = The subject in which the attribute is deleted \newline
$attr$ = The deleted attribute \\ \hline

SPARQL used for detection &
\vspace{-0.4cm}
\begin{verbatim}
SELECT ?x ?attr WHERE { 
GRAPH <v2> {  
FILTER NOT EXISTS {  
{?attr rdf:type 
qb:DimensionProperty.} 
UNION  
{?attr rdf:type 
qb:MeasureProperty.} 
UNION 
{?attr rdf:type 
qb:CodedProperty.} 
} 
?x qb:attribute ?attr. 
} 
FILTER NOT EXISTS { GRAPH <v1> { 
?x qb:attribute ?attr. 
}
}
}
\end{verbatim}
\vspace{-0.4cm} &
\vspace{-0.4cm}
\begin{verbatim}
SELECT ?x ?attr WHERE { 
GRAPH <v1> {  
FILTER NOT EXISTS {  
{?attr rdf:type 
qb:DimensionProperty.} 
UNION  
{?attr rdf:type 
qb:MeasureProperty.} 
UNION 
{?attr rdf:type 
qb:CodedProperty.} 
} 
?x qb:attribute ?attr. 
} 
FILTER NOT EXISTS { GRAPH <v2> { 
?x qb:attribute ?attr. 
}
}
}
\end{verbatim}
\vspace{-0.4cm}  \\ \hline

$\chng^+$ &
\hasAttribute{x}{attr}
 &
$\emptyset$ \\ \hline

$\chng^-$ &
$\emptyset$  &
\hasAttribute{x}{attr}
 \\ \hline

$\cond_{old}$ &
--  &
\begin{verbatim}
FILTER NOT EXISTS {  
{?attr rdf:type 
qb:DimensionProperty.} 
UNION  
{?attr rdf:type 
qb:MeasureProperty.} 
UNION 
{?attr rdf:type 
qb:CodedProperty.} 
} 
?x qb:attribute ?attr. }
\end{verbatim}  \\ \hline
$\cond_{new}$ &
\begin{verbatim}
FILTER NOT EXISTS {  
{?attr rdf:type 
qb:DimensionProperty.} 
UNION  
{?attr rdf:type 
qb:MeasureProperty.} 
UNION 
{?attr rdf:type 
qb:CodedProperty.} 
} 
?x qb:attribute ?attr. }
\end{verbatim}  &
-- \\ \hline

\end{tabular}
\end{footnotesize}
\newline
\end{table*}


\begin{table*}[h]
\begin{footnotesize}
\begin{tabular}{|p{2cm}|p{6.5cm}|p{6.5cm}|}
\hline
Change &
\textbf{\op{Add\_Generic\_Value\_to\_Observation($o,p,v$)}} &
\textbf{\op{Delete\_Generic\_Value\_from\_Observation($o,p,v$)}} \\ \hline

Intuition &
Add a generic value to observation &
Delete a generic value from an observation \\ \hline

Parameters &
$o$ = The observation \newline
$p$ = The property related as predicate of the observation \newline
$v$ = The added value
&
$o$ = The observation \newline
$p$ = The property related as predicate of the observation \newline
$v$ = The deleted value
\\ \hline

SPARQL used for detection &
\vspace{-0.4cm}
\begin{verbatim}
SELECT ?o ?p ?v WHERE { 
GRAPH <v2> { 
FILTER NOT EXISTS { 
{?p rdf:type qb:DimensionProperty.} 
UNION 
{?p rdf:type qb:MeasureProperty.} 
} 
} 
GRAPH <v2> { 
?o qb:dataSet ?ds. 
?ds qb:structure ?ft. 
?ft qb:component ?cs. 
?cs qb:componentProperty ?p. 
?p rdfs:range ?v. 
} 
FILTER NOT EXISTS { GRAPH <v1> { 
?o qb:dataSet ?ds. 
?ds qb:structure ?ft. 
?ft qb:component ?cs. 
?cs qb:componentProperty ?p. 
?p rdfs:range ?v.  
}
}
}
\end{verbatim}
\vspace{-0.4cm} &
\vspace{-0.4cm}
\begin{verbatim}
SELECT ?o ?p ?v WHERE { 
GRAPH <v1> { 
FILTER NOT EXISTS { 
{?p rdf:type qb:DimensionProperty.} 
UNION 
{?p rdf:type qb:MeasureProperty.} 
} 
} 
GRAPH <v1> { 
?o qb:dataSet ?ds. 
?ds qb:structure ?ft. 
?ft qb:component ?cs. 
?cs qb:componentProperty ?p. 
?p rdfs:range ?v. 
} 
FILTER NOT EXISTS { GRAPH <v2> { 
?o qb:dataSet ?ds. 
?ds qb:structure ?ft. 
?ft qb:component ?cs. 
?cs qb:componentProperty ?p. 
?p rdfs:range ?v.  
}
}
}
\end{verbatim}
\vspace{-0.4cm}  \\ \hline

$\chng^+$ &
\hasDataset{o}{ds},
\hasStructure{ds}{ft},
\hasComponent{ft}{cs},
\hasComponentProperty{cs}{p},
\hasRange{p}{v}
&
$\emptyset$ \\ \hline

$\chng^-$ &
$\emptyset$  &
\hasDataset{o}{ds},
\hasStructure{ds}{ft},
\hasComponent{ft}{cs},
\hasComponentProperty{cs}{p},
\hasRange{p}{v}
\\ \hline

$\cond_{old}$ &
--  &
\begin{verbatim}
FILTER NOT EXISTS { 
{?p rdf:type qb:DimensionProperty.} 
UNION 
{?p rdf:type qb:MeasureProperty.} 
}
\end{verbatim}  \\ \hline
$\cond_{new}$ &
\begin{verbatim}
FILTER NOT EXISTS { 
{?p rdf:type qb:DimensionProperty.} 
UNION 
{?p rdf:type qb:MeasureProperty.} 
}
\end{verbatim}  &
-- \\ \hline

\end{tabular}
\end{footnotesize}
\newline
\end{table*}

}
{}



\end{document}